\documentclass[12pt]{article}

\title{On the numerical treatment of dissipative particle dynamics and related systems}

\author{Benedict Leimkuhler and Xiaocheng Shang\footnote{Corresponding author. Email: \href{mailto:x.shang@ed.ac.uk}{x.shang@ed.ac.uk}} \\
\small{School of Mathematics, University of Edinburgh, Edinburgh, EH9 3JZ, UK} }

\date{\today}

\usepackage{hyperref}
\usepackage{graphicx}

\usepackage{epstopdf} 

\usepackage{caption}
\DeclareGraphicsExtensions{.eps,.mps,.pdf,.jpg,.png}
\graphicspath{{figures/}{../figures/}}

\usepackage{fancyhdr}

\usepackage{amsmath}
\usepackage{amsfonts}

\usepackage{dsfont}

\usepackage[titletoc,toc,title]{appendix}

\usepackage[margin=3.0cm]{geometry}

\newcommand{\dd}{{\rm d}}

\usepackage{array}
\newcolumntype{C}[1]{>{\centering\let\newline\\\arraybackslash\hspace{0pt}}m{#1}}

\begin{document}

\maketitle

\begin{abstract}
We review and compare numerical methods that simultaneously control temperature while preserving the momentum, a family of particle simulation methods commonly used for the modelling of complex fluids and polymers.   The class of methods considered includes dissipative particle dynamics (DPD) as well as extended stochastic-dynamics models incorporating a generalized pairwise thermostat scheme in which stochastic forces are eliminated and the coefficient of dissipation is treated as an additional auxiliary variable subject to a feedback (kinetic energy) control mechanism.  In the latter case, we consider the addition of a coupling of the auxiliary variable, as in the Nos\'{e}-Hoover-Langevin (NHL) method, with  stochastic dynamics to ensure ergodicity, and find that the convergence of ensemble averages is substantially improved.   To this end, splitting methods are developed and studied in terms of their thermodynamic accuracy, two-point correlation functions, and convergence.   In terms of computational efficiency as measured by the ratio of thermodynamic accuracy to CPU time, we report significant advantages in simulation for the pairwise NHL method compared to popular alternative schemes (up to an 80\% improvement), without degradation of convergence rate.   The momentum-conserving thermostat technique described here provides a consistent hydrodynamic model in the low-friction regime, but it will also be of use in both equilibrium and nonequilibrium molecular simulation applications owing to its efficiency and simple numerical implementation.
\end{abstract}




\section{Introduction}

Stochastic momentum-conserving thermostats, which correctly capture long-ranged hydrodynamic interactions, are increasingly popular tools for simulation of simple and complex fluids \cite{Allen2007}.   The first important scheme of this type was dissipative particle dynamics (DPD), introduced by Hoogerbrugge and Koelman \cite{Hoogerbrugge1992} in 1992 for simulating complex hydrodynamic behavior at a mesoscopic level that is not accessible by conventional molecular dynamics (MD) \cite{Allen1989,Frenkel2001}.  In DPD, a collection of fluid molecules are grouped at the coarse-grained level and treated as a discrete particle.  These particles interact at short range in a soft potential, thereby allowing larger timesteps than would be possible in MD, while simultaneously decreasing the number of degrees of freedom required.   DPD thus bridges the gap between microscale (atomistic methods, e.g. molecular dynamics) and macroscale (continuum methods, e.g. Navier-Stokes) models and can be used to recover thermodynamic, dynamical and rheological properties of complex fluids, with applications to colloidal particles \cite{Boek1997}, polymer molecules \cite{Schlijper1995} and fluid mixtures \cite{Coveney1996}.

A great deal of effort has been devoted to the design of simple, efficient and accurate numerical methods to solve the DPD system due to its promising prospects from the applications perspective. In the example of lipid bilayers, new phenomena arise as the time scale of the system that we are investigating is increased \cite{Groot2004}.  However, not all algorithms are rigorously founded and may not perform satisfactorily in large scale simulations (see discussions in \cite{Besold2000,Vattulainen2002,Nikunen2003,Chaudhri2010}).    All the methods exhibit pronounced artifacts with increasing integration stepsizes due to the discretization error, typically manifest in the form of statistical bias in the calculation of thermodynamic averages.   A previous study \cite{Jakobsen2005} suggested that, without performing serious checks for each method, the only reliable approach is to use vanishingly small stepsizes, since most of the schemes proposed are formally convergent at some order of accuracy.    However, we argue that using very small stepsize significantly limits the time scales accessible for DPD simulation, particularly in large scale simulations.

One of the key questions we tackle in this article is ``What is the largest integration stepsize that can be used without damaging both static and dynamical properties?''  Answering this question leads to a better understanding of the overall efficiency of each method and the validity of the schemes in the computation-intensive setting (large particle number/long-time interval).  Recently, a systematic approach to thermodynamic bias in numerical computations has been used to study the accuracy and efficiency of methods for Langevin dynamics \cite{Bou-Rabee2010,Leimkuhler2013,Leimkuhler2013a}.   The approach suggested is to determine the order of accuracy of a stochastic scheme in relation to its effective invariant distribution (and, thus, with respect to steady state averages computed using numerical trajectories).  This technique has led to greatly improved numerical methods for Langevin dynamics, and it is in principle applicable to momentum-conserving thermostat schemes as well.  However these results are based on asymptotic expansion and thus are only relevant in the limit of small stepsize (whereas we are interested in the large stepsize threshold).  Moreover we find that the analytical computations necessary to perform expansions in the DPD and pairwise NHL cases are in typical cases highly complex;  we therefore restrict ourselves in this article to outlining some fundamental and  illustrative applications of the theory.  In particular for certain symmetric methods, we can demonstrate the even-order approximation of long-time averages.    The same conclusion may be reached for additional schemes of a specific structure (related to the Geometric Langevin Algorithms of \cite{Bou-Rabee2010}).

A new stochastic momentum-conserving thermostat is introduced in this article that can be used in place of DPD in the low-friction regime or in nonequilibrium molecular dynamics (NEMD) based on stochastic extension of a scheme in a recent paper by Allen and Schmid \cite{Allen2007}.   This method is particularly inexpensive to implement and is found to have a very high stepsize stability threshold compared to alternatives.

In typical cases, for understanding the stepsize stability threshold and the performance of different schemes, we are forced to rely on numerical experiment.  An excellent survey of the performance of a number of methods was undertaken by Nikunen \emph{et al.} \cite{Nikunen2003} in 2003. Since then, and despite many additions to the arsenal of methods, such a comparison has been lacking.  To this end, we test a number of popular methods \cite{Besold2000,Shardlow2003,DeFabritiis2006} from the DPD literature together with additional methods \cite{Lowe1999,Peters2004,Stoyanov2005} that are used in popular software packages.   For each method, we examine calculations such as kinetic and configurational temperatures, average potential energy, radial distribution function, velocity autocorrelation function and transverse momentum autocorrelation function, which gives information on the rotational relaxation process.

The outline of the article is as follows. We first review the formulation and various integration schemes of DPD in Section \ref{Section:DPD}. Two extended variable momentum-conserving thermostats are presented in Section \ref{Section:Extended_Variable} with the latter newly proposed in this article. In Section \ref{Section:Error_in_DPD}, we demonstrate the definitions of error in DPD simulation and summarize the orders of accuracy both theoretically and numerically in a number of methods. Various numerical experiments are carried out in Section \ref{Section:Numerical_Experiments} to compare all the schemes described in the article. We summarize our findings in Section \ref{Section:Conclusions}.

\section{Dissipative Particle Dynamics}
\label{Section:DPD}

\subsection{Formulation of DPD}

The original version of DPD \cite{Hoogerbrugge1992} updated the system in discrete time steps and was later  reformulated by Espa\~{n}ol and Warren in 1995 \cite{Espanol1995} as a system of stochastic differential equations (SDEs).

Consider $N$ particles with positions $\mathbf{q}_{i}$, momenta $\mathbf{p}_{i}$ and masses $m_{i}$ for $i=1,...,N$ evolving in dimension $d$. The time evolution of the DPD particles, each of which represents a cluster of molecules,
is governed by Newton's equations
\begin{equation}\label{eqn:Newton}
  \frac{\dd \mathbf{q}_{i}}{\dd t}=\frac{\mathbf{p}_{i}}{m_{i}}, \quad \frac{\dd \mathbf{p}_{i}}{\dd t}=\mathbf{F}_{i},
\end{equation}
where $\mathbf{F}_{i}$ is the total interparticle force acting on particle $i$ due to the presence of the other particles. The force is composed of three pairwise contributions
\begin{equation}\label{eqn:Forces}
  \mathbf{F}_{i}=\sum_{j\neq i}(\mathbf{F}_{ij}^{C}+\mathbf{F}_{ij}^{D}+\mathbf{F}_{ij}^{R}),
\end{equation}
where $\mathbf{F}_{ij}^{C}$, $\mathbf{F}_{ij}^{D}$ and $\mathbf{F}_{ij}^{R}$ represent conservative, dissipative and random forces, acting on particle $i$ due to the $j$-th particle, respectively.

The conservative force controlling the thermodynamics of the DPD system is normally \cite{Groot1997} chosen as
\begin{equation}\label{eqn:Conservative_Force}
  \mathbf{F}_{ij}^{C}=
  \begin{cases}
  a_{ij}\biggl(1-\displaystyle\frac{q_{ij}}{r_{c}}\biggr)\hat{\mathbf{q}}_{ij}, & q_{ij}<r_{c};\\
  0, & q_{ij}\geq r_{c},
  \end{cases}
\end{equation}
where parameters $a_{ij}$ are symmetric $(a_{ij}=a_{ji}\geq0)$ representing the maximum repulsion strength between particle $i$ and particle $j$, and $r_{c}$ is a certain cutoff radius. The relative positions are denoted by $\mathbf{q}_{ij}=\mathbf{q}_{i}-\mathbf{q}_{j}$ with length $q_{ij}=|\mathbf{q}_{ij}|$ and the unit direction from $\mathbf{q}_{j}$ to $\mathbf{q}_{i}$ by $\hat{\mathbf{q}}_{ij}=\mathbf{q}_{ij}/q_{ij}$.  It should be noted that there are many alternative choices for the mesoscopic force field.  For example, a Lennard-Jones type potential has been used in combination with DPD thermostat in simulating equilibrium and nonequilibrium molecular dynamics \cite{Soddemann2003}.  Potentials may also be obtained by direct coarse-graining of a microscopic system \cite{Lyubartsev1995,Lei2010}.  Although we assume the simple force law (\ref{eqn:Conservative_Force}), many of the observations of this article would hold for more general conservative force fields.

The dissipative and random forces behave as a thermostat to keep the temperature constant and are given by
\begin{align}
  \mathbf{F}_{ij}^{D} &= -\gamma \omega^{D}(q_{ij})(\hat{\mathbf{q}}_{ij}\cdot \mathbf{v}_{ij})\hat{\mathbf{q}}_{ij}, \\
  \mathbf{F}_{ij}^{R} &= \sigma \omega^{R}(q_{ij})\theta_{ij}\hat{\mathbf{q}}_{ij},
\end{align}
where $\gamma$ and $\sigma$ represent the dissipative and random strengths respectively, $\omega^{D}$ and $\omega^{R}$ are position-dependent weight functions, and the relative velocities are denoted by $\mathbf{v}_{ij}=\mathbf{p}_{i}/m_{i}-\mathbf{p}_{j}/m_{j}$. $\theta_{ij}$  is a symmetric ($\theta_{ij}=\theta_{ji}$) Gaussian white-noise term with the following stochastic property
\begin{equation}
  \langle\theta_{ij}(t)\rangle=0, \quad \quad \langle\theta_{ij}(t)\theta_{kl}(t')\rangle=(\delta_{ik}\delta_{jl} + \delta_{il}\delta_{jk})\delta(t-t'),
\end{equation}
and is chosen independently for each pair of interacting particles at each time step.

The canonical ensemble was not preserved in the original formulation of DPD of \cite{Hoogerbrugge1992}.  This has been corrected by Espa\~{n}ol and Warren \cite{Espanol1995} who showed that certain conditions have to be satisfied to guarantee that the DPD system has the same invariant distribution as would be obtained in cases where the dissipative and random forces are put to zero, namely:
\begin{equation}\label{eqn:Fluctuation_Dissipative_Theorem}
  \omega^{D}(q_{ij})=[\omega^{R}(q_{ij})]^{2}, \quad \sigma^{2}=2\gamma k_{B}T,
\end{equation}
where $k_{B}$ is the Boltzmann constant and $T$ the equilibrium temperature. This is the fluctuation-dissipation theorem for the DPD thermostat involving dissipative and random forces only. Then it can be easily shown that the canonical ensemble is preserved with invariant distribution defined by the density
\begin{equation}\label{eqn:Invariant_Distribution_DPD}
  \rho_{\beta}(\mathbf{q},\mathbf{p})=\frac{1}{Z}\exp \left(-\beta H(\mathbf{q},\mathbf{p}) \right),
\end{equation}
where $\beta^{-1}=k_{B}T$, $Z$ is the partition function and $H(\mathbf{q},\mathbf{p})$ is the Hamiltonian defined as
\begin{equation}
  H(\mathbf{q},\mathbf{p}) = \displaystyle \sum_{i} \frac{\mathbf{p}_{i}^{2}}{2m_{i}} + U(\mathbf{q}) = \displaystyle \sum_{i} \frac{\mathbf{p}_{i}^{2}}{2m_{i}} + \frac{1}{2}\sum_{i} \sum_{j\neq i}U(q_{ij}),
\end{equation}
where the ``soft'' pair potential energy $U(q_{ij})$ corresponding to the conservative force (\ref{eqn:Conservative_Force}) is defined as
\begin{equation}\label{eqn:Soft_Potential}
  U(q_{ij})=
  \begin{cases}
    \displaystyle\frac{a_{ij}r_{c}}{2}\biggl(1-\displaystyle\frac{q_{ij}}{r_{c}}\biggr)^{2}, & q_{ij}<r_{c};\\
    0, & q_{ij}\geq r_{c}.
  \end{cases}
\end{equation}
Although we write the density (\ref{eqn:Invariant_Distribution_DPD}) as an exponential, we note that if the total momentum is conserved, the density should be replaced by
\begin{equation*}
  \rho_{\beta}(\mathbf{q},\mathbf{p})=\frac{1}{Z} \exp(-\beta H(\mathbf{q},\mathbf{p})) \times \delta\left[ \sum_i p_{i,x}- \pi_x \right] \delta\left [\sum_i p_{i,y} - \pi_y \right] \delta\left [\sum_i p_{i,z} - \pi_z \right],
\end{equation*}
where $\boldsymbol{\pi}=(\pi_x,\pi_y,\pi_z)$ is the total momentum vector. A similar modification would be needed if the angular momentum were also conserved. It is worth pointing out that the ergodicity of the DPD system has only been demonstrated in the case of high particle density in one dimension by Shardlow and Yan \cite{Shardlow2006}.

Due to the fact that the algorithm depends on relative velocities and the interactions between particles are symmetric, both total and angular momenta are conserved, DPD is therefore an isotropic Galilean-invariant thermostat which preserves hydrodynamics \cite{Allen2007,Moeendarbary2009}.  If periodic boundary conditions are used, the angular momentum will be destroyed as a conserved quantity, but the symmetric character of the force field has important consequences e.g. for the relaxation of a localized rotational excitation. An early study of constant-temperature molecular dynamics with momentum conservation can be found in \cite{Cho1993}, but the techniques discussed there are not that relevant to the present work.

One of the two weight functions can be chosen arbitrarily without changing thermodynamic equilibrium. A simple and widely-used choice reads
\begin{equation}\label{eqn:Wight_Function_R}
  \omega^{R}(q_{ij})=
  \begin{cases}
  1-\displaystyle\frac{q_{ij}}{r_{c}}, & q_{ij}<r_{c};\\
  0, & q_{ij}\geq r_{c},
  \end{cases}
\end{equation}
in which case the conservative force (\ref{eqn:Conservative_Force}) can be written in a compact way $\mathbf{F}_{ij}^{C}=a_{ij}\omega^{R}(q_{ij})\hat{\mathbf{q}}_{ij}$.

To make the presentation simpler, a compact form of the SDEs of the DPD system (\ref{eqn:Newton}) (for particle $i$) may be used \cite{Allen2007}
\begin{equation}
  \label{eqn:DPD_system}
  \begin{aligned}
    \dd \mathbf{q}_{i} &= m_{i}^{-1}\mathbf{p}_{i}\dd t, \\
    \dd \mathbf{p}_{i} &= \mathbf{F}^{C}_{i}(\mathbf{q})\dd t - \gamma \mathbf{V}_{i}(\mathbf{q},\mathbf{p})\dd t + \sigma \mathbf{R}_{i}(\mathbf{q},t),
  \end{aligned}
\end{equation}
where $\mathbf{F}^{C}_{i}(\mathbf{q})$ is the total conservative forces acting on particle $i$
\begin{equation}
  \mathbf{F}^{C}_{i}(\mathbf{q}) = \sum_{j\neq i}\mathbf{F}_{ij}^{C}(q_{ij}) = -\nabla_{\mathbf{q}_{i}}U(\mathbf{q}),
\end{equation}
and $\mathbf{V}_{i}(\mathbf{q},\mathbf{p})$ and $\mathbf{R}_{i}(\mathbf{q},\mathbf{p},t)$ are defined respectively as
\begin{align}
  \mathbf{V}_{i}(\mathbf{q},\mathbf{p})   &= \sum_{j\neq i}\omega^{D}(q_{ij})(\hat{\mathbf{q}}_{ij}\cdot \mathbf{v}_{ij})\hat{\mathbf{q}}_{ij}, \\
  \mathbf{R}_{i}(\mathbf{q},t) &= \sum_{j\neq i}\omega^{R}(q_{ij})\hat{\mathbf{q}}_{ij}\dd \mathrm{W}_{ij}(t),
\end{align}
where $\dd \mathrm{W}_{ij}(t)=\dd \mathrm{W}_{ji}(t)$ are independent increments of a Wiener process with mean zero and variance $\dd t$ \cite{Groot1997}.

\subsection{Numerical Integration Schemes}
\label{Subsection:Numerical_Integration_Schemes}

Due to the soft repulsive potential (\ref{eqn:Soft_Potential}), the major advantage of the DPD method is that the stepsize used in simulations may be much larger than those of conventional MD simulations with Lennard-Jones internuclear potentials for instance.   This feature is crucial especially when a very long simulation time is required. However, large stepsizes may result in errors in computed thermodynamic quantities.  Whereas for molecular dynamics, simulations are performed at or near the stability threshold defined by stiff components such as harmonic bonds \cite{Leimkuhler2013a}, DPD simulations may be perfectly stable over a wide range of stepsizes for which errors in averages are very large.     There have been many attempts to develop accurate and efficient numerical methods that allow larger stepsizes.  This is currently an active field of research.

In the early days of DPD, a number of integration schemes were proposed based on the well-recognized velocity Verlet scheme \cite{Verlet1967} widely used in classical MD simulations \cite{Allen1989,Frenkel2001}.   Specific examples are the integrator of Groot and Warren \cite{Groot1997} (GW), the method of Gibson \emph{et al.} \cite{Gibson1999} (GCC) and the DPD velocity-Verlet integrator, which refers to as DPD-VV, of Besold \emph{et al.} \cite{Besold2000} (see more discussions in \cite{Vattulainen2002}). Both the GW and GCC integrators incorporate a parameter $\lambda$, which has to be chosen carefully for specific model parameters, to reduce unphysical artifacts. Nevertheless, as reported in \cite{Besold2000,Vattulainen2002}, all the integrators mentioned above display pronounced artifacts, especially when the stepsize is large, due to the fact that the velocities and dissipative forces depend on each other implicitly and thus need to be updated in a self-consistent fashion.

In the spirit of the self-consistent leap-frog integrator introduced by Pagonabarraga \emph{et al.} \cite{Pagonabarraga1998}, Besold \emph{et al.} \cite{Besold2000} proposed a self-consistent velocity Verlet scheme, which we label SC-VV, in which, at the end of each iteration step, the velocity is ``corrected'' based on the newly-calculated dissipative force until the deviation between the instantaneous kinetic temperature and the target temperature is less than a certain value of ``tolerance''. It should be noted that there is no such ``correction'' in the DPD-VV scheme but still the dissipative force is recalculated once, using the up-to-date velocities (momenta).  A variant of the SC-VV scheme, SC-Th,  introduced in \cite{Besold2000}, couples the original DPD system to an auxiliary Nos\'{e}-Hoover thermostat \cite{Nose1984,Nose1984a,Hoover1985,Hoover1991} to provide direct kinetic temperature control. Overall, the self-consistent schemes do reduce unphysical artifacts to some extent, however, it is also well-documented \cite{Nikunen2003,Chaudhri2010} that they can be substantially slower than standard methods, depending how small the tolerance is. Therefore, computationally expensive self-consistent methods are not the choices present in typical software packages.

Although it has been demonstrated that relatively small stepsizes must be used in the DPD-VV scheme to produce correct static and dynamical properties \cite{Vattulainen2002,Nikunen2003}, the DPD-VV scheme remains one of the most popular methods for DPD system in software packages due to its efficiency and ease of implementation (particularly in parallel computing). For this reason, the DPD-VV method has been chosen as the ``benchmark'' to compare with other schemes in this article.

Several novel integration schemes have been proposed over the years, such as the approach by den Otter and Clarke \cite{DenOtter2001}, the extended DPD by Cotter and Reich \cite{Cotter2003}, the multiple time step schemes by Jakobsen \emph{et al.} \cite{Jakobsen2006}, the Trotter-splitting methods by Thalmann and Farago \cite{Thalmann2007} and most recently the algorithm by Goga \emph{et al.} \cite{Goga2012}. In this article, we focus on schemes that have been included in popular software packages (i.e. DPD-VV \cite{Besold2000}, Peters thermostat \cite{Peters2004}, Lowe-Andersen thermostat \cite{Lowe1999} and Nos\'{e}-Hoover-Lowe-Andersen thermostat \cite{Stoyanov2005} in the DL\_MESO \cite{Seaton2013} package) with two promising splitting methods by Shardlow \cite{Shardlow2003} and De Fabritiis \emph{et al.} \cite{DeFabritiis2006}, respectively. Particularly, Nikunen \emph{et al.} \cite{Nikunen2003} in 2003 showed that the performances of Lowe-Andersen thermostat and Shardlow's scheme are superior to those of several other schemes for a number of different observables. Recently, Shardlow-like splitting algorithms have been further applied in DPD with various fixed conditions \cite{Lisal2011}, and its parallel implementation has also been developed in \cite{Larentzos2014}.

The full details of the integration steps of each method described in this article are presented in a common language in Appendix \ref{Section:Appendix_Schemes}.

\subsubsection{DPD Velocity-Verlet}

For integration stepsize $h$, the simple DPD-VV integrator \cite{Besold2000} reads
\begin{align}
  \mathbf{p}_{i}^{n+1/2} &= \mathbf{p}_{i}^{n} + \left( h\mathbf{F}^{C}_{i}(\mathbf{q}^{n}) + h\mathbf{F}^{D}_{i}(\mathbf{q}^{n},\mathbf{p}^{n}) + \sqrt{h}\mathbf{F}^{R}_{i}(\mathbf{q}^{n}) \right) /2,  \label{eqn:DPD_VV_1} \\
  \mathbf{q}_{i}^{n+1} &=  \mathbf{q}_{i}^{n} + hm_{i}^{-1}\mathbf{p}_{i}^{n+1/2} , \label{eqn:DPD_VV_2} \\
  \mathbf{p}_{i}^{n+1} &= \mathbf{p}_{i}^{n+1/2} + \left( h\mathbf{F}^{C}_{i}(\mathbf{q}^{n+1}) + h\mathbf{F}^{D}_{i}(\mathbf{q}^{n+1},\mathbf{p}^{n+1/2}) + \sqrt{h}\mathbf{F}^{R}_{i}(\mathbf{q}^{n+1}) \right) /2.
\end{align}
In fact, the DPD-VV scheme has two differences compared to the standard velocity Verlet method \cite{Verlet1967}: (1) the forces are not just the conventional conservative forces, but include dissipative and random forces, as well; (2) the dissipative forces have to be updated for a second time at the end of each integration step by using the up-to-date velocities (momenta), $\mathbf{F}^{D}_{i}(\mathbf{q}^{n+1},\mathbf{p}^{n+1})$, with the first update taking place right after the positions are updated at each step (see more details in Appendix \ref{Section:Appendix_Schemes}). It has been shown that the performance of the DPD-VV method would be significantly improved \cite{Besold2000,Vattulainen2002} by simply doing the additional update of the dissipative forces in each step, which is actually not time-consuming if one makes use of computation-saving devices such as Verlet neighbor lists \cite{Verlet1967}. Note that both the GW integrator \cite{Groot1997} of Groot and Warren and the modified Verlet method mentioned by Shardlow in \cite{Shardlow2003} do not incorporate the additional update.

It is important to observe that, unlike the standard velocity Verlet method (second order), the DPD-VV scheme should only give first order convergence for the invariant measure (see more details in Section \ref{Section:Error_in_DPD}) due to the fact that the momentum is not updated in a symmetric manner. However, second order convergence for averages was observed in the numerical experiments in Section \ref{Section:Numerical_Experiments}. Moreover, the term $\sqrt{h}/2$ multiplying the random forces, which would be expected to be different in this type of splitting of Langevin dynamics, must be present when random forces are reused in the subsequent step; it ensures that the diffusion coefficient of the particles is independent of the integration timestep (see \cite{Groot1997} for further discussion).

\subsubsection{Shardlow's Splitting Method}

Splitting techniques were studied by Shardlow \cite{Shardlow2003} based on dividing the vector field of the DPD system into three parts, the first two of which represent the vector field of the Hamiltonian system associated with kinetic and potential energies, and the last term is the remaining Langevin equation (actually Ornstein-Uhlenbeck process with positions fixed) involving the dissipative and random forces. Two integrators, termed there S1 and S2, have been proposed for treating this system in \cite{Shardlow2003}.    Only the S1 method will be examined here as it is more efficient than S2.   This scheme relies on the method of Br\"{u}nger, Brooks and Karplus (BBK) \cite{Brunger1984} to solve the Langevin part, followed by the standard velocity Verlet scheme for the conservative part.

In describing Shardlow's method or another splitting scheme, we use the formal notation of the generator of the diffusion as in, for example, \cite{DeFabritiis2006},  \cite{Serrano2006},  and\cite{Thalmann2007}.

We first separate the system of stochastic differential equations for DPD (\ref{eqn:DPD_system}) into three pieces, which we label as A, B and O:
\begin{equation}
  \dd \left[ \begin{array}{c} \mathbf{q}_{i} \\ \mathbf{p}_{i} \end{array} \right] = \underbrace{\left[ \begin{array}{c} m_{i}^{-1}\mathbf{p}_{i} \\ \mathbf{0} \end{array} \right] \dd t}_\text{A} + \underbrace{\left[ \begin{array}{c} \mathbf{0} \\ \mathbf{F}^{C}_{i} \end{array} \right] \dd t}_\text{B} + \underbrace{\left[ \begin{array}{c} \mathbf{0} \\ - \gamma \mathbf{V}_{i}\dd t + \sigma \mathbf{R}_{i} \end{array} \right]}_\text{O}.
\end{equation}
The generators for each part of the SDE may be written out as follows:
\begin{align}
  \mathcal{L}_\text{A} &= \sum_{i}\frac{\mathbf{p}_{i}}{m_{i}} \cdot \nabla_{\mathbf{q}_{i}}, \label{eqn:generator_A} \\
  \mathcal{L}_\text{B} &= \sum_{i} \mathbf{F}_{i}^{C} \cdot \nabla_{\mathbf{p}_{i}} = -\sum_{i} \nabla_{\mathbf{q}_{i}}U(\mathbf{q}) \cdot \nabla_{\mathbf{p}_{i}}, \label{eqn:generator_B} \\
  \mathcal{L}_\text{O} &= \sum_{i} \sum_{j\neq i}  \left( - \gamma \omega^{D}(q_{ij})(\hat{\mathbf{q}}_{ij}\cdot \mathbf{v}_{ij}) + \displaystyle\frac{\sigma^{2}}{2}[\omega^{R}(q_{ij})]^{2} \hat{\mathbf{q}}_{ij} \cdot \left( \nabla_{\mathbf{p}_{i}} - \nabla_{\mathbf{p}_{j}} \right) \right) \hat{\mathbf{q}}_{ij} \cdot \nabla_{\mathbf{p}_{i}}. \label{eqn:generator_DPD_O}
\end{align}
Thus, the generator for the DPD system can be written as
\begin{equation}\label{eqn:generator_DPD}
  \mathcal{L}_\text{DPD} = \mathcal{L}_\text{A} + \mathcal{L}_\text{B} + \mathcal{L}_\text{O}.
\end{equation}

The flow map (or phase space propagator) of the system may be written in the shorthand notation
\begin{equation*}
{\cal F}_t = e^{t \mathcal{L}_{\text{DPD}}},
\end{equation*}
where the exponential map is here used formally to denote the solution operator.  Approximations of ${\cal F}_t$ may then be obtained as products (taken in different arrangements) of exponentials of the various terms of the splitting.   For example,  the phase space propagation of Shardlow's S1 splitting method \cite{Shardlow2003}, termed as DPD-S1, can be written as
\begin{equation}\label{eqn:DPD_S1_propagator}
\exp\left(h\hat{\mathcal{L}}_\text{DPD-S1} \right)
= \exp\left(h\mathcal{L}_\text{O}\right) \exp\left(\frac{h}{2}\mathcal{L}_\text{B}\right) \exp\left(h\mathcal{L}_\text{A}\right) \exp\left(\frac{h}{2}\mathcal{L}_\text{B}\right),
\end{equation}
where $h$ is the stepsize and $\exp\left(h\mathcal{L}_f\right)$ represents the phase space propagator associated with the corresponding vector field $f$.   In Shardlow's approach,  the vector field O is further split into each interacting pair. Therefore, the propagation of the O part is broken down into many terms:
\begin{equation*}
  \exp\left(h\hat{\mathcal{L}}_\text{O}\right) = \exp\left(h\mathcal{L}_{\text{O}_{N-1,N}}\right) \cdots \exp\left(h\mathcal{L}_{\text{O}_{1,3}}\right) \exp\left(h\mathcal{L}_{\text{O}_{1,2}}\right),
\end{equation*}
where the operators associated with each interacting pair are defined as
\begin{equation}
  \mathcal{L}_{\text{O}_{i,j}} = \left( - \gamma \omega^{D}(q_{ij})(\hat{\mathbf{q}}_{ij}\cdot \mathbf{v}_{ij}) + \displaystyle\frac{\sigma^{2}}{2}[\omega^{R}(q_{ij})]^{2} \hat{\mathbf{q}}_{ij} \cdot \left( \nabla_{\mathbf{p}_{i}} - \nabla_{\mathbf{p}_{j}} \right) \right) \hat{\mathbf{q}}_{ij} \cdot \nabla_{\mathbf{p}_{i}}.
\end{equation}
Each interacting pair preserves the invariant distribution $\rho_{\beta}$ (\ref{eqn:Invariant_Distribution_DPD}).   As a shorthand, we may term the DPD-S1 method  OBAB (similarly, the S2 method of Shardlow would be equivalent to BABOBAB in the same language), where, in both cases, it is to be understood that the steplengths associated with various operations are uniform and span the interval $h$.   Thus the B step in OBAB is taken with a step length of $h/2$, while O with a steplength of $h$.

\subsubsection{DPD-Trotter Scheme}
\label{Subsubsection:DPD_Trotter}

The Trotter formula \cite{Strang1968} that has been widely used in molecular simulations was investigated and further applied to split the DPD generator (\ref{eqn:generator_DPD}) in an ``optimal'' way to reduce artifacts and maintain good temperature control \cite{DeFabritiis2006,Serrano2006}. A new scheme, referred to as DPD-Trotter, was introduced but few numerical simulations have been presented and therefore we incorporate it in the comparisons.

In the stochastic DPD-Trotter scheme, the standard DPD system (\ref{eqn:DPD_system}) is split into two parts, which are labeled ``A'' and ``S'' as indicated below
\begin{equation}
  \dd \left[ \begin{array}{c} \mathbf{q}_{i} \\ \mathbf{p}_{i} \end{array} \right] = \underbrace{\left[ \begin{array}{c} m_{i}^{-1}\mathbf{p}_{i} \\ \mathbf{0} \end{array} \right] \dd t}_\text{A} + \underbrace{\left[ \begin{array}{c} \mathbf{0} \\ \mathbf{F}^{C}_{i}\dd t - \gamma \mathbf{V}_{i}\dd t + \sigma \mathbf{R}_{i} \end{array} \right]}_\text{S}.
\end{equation}
The corresponding operator of part A is exactly the same as in Shardlow's method, while the operator of part S is actually the sum of the operators of B and O defined above
\begin{equation}
  \mathcal{L}_\text{S} = \mathcal{L}_\text{B} + \mathcal{L}_\text{O}.
\end{equation}
As in Shardlow's method,  the vector field S is further split into each interacting pair; these pair interactions are exactly solvable (in the sense of distributional fidelity). In fact, for $j>i$, subtracting $\dd \mathbf{v}_{j}$ from $\dd \mathbf{v}_{i}$ and multiplying $\hat{\mathbf{q}}_{ij}$ on both sides gives
\begin{equation}\label{eqn:Trotter_exact}
    m_{ij}\dd v_{ij} = F_{ij}^{C}(q_{ij})\dd t - \gamma \omega^{D}(q_{ij})v_{ij}\dd t + \sigma \omega^{R}(q_{ij})\dd \mathrm{W}_{ij},
\end{equation}
where $m_{ij}=m_{i}m_{j}/(m_{i}+m_{j})$ is the ``reduced mass'', $v_{ij}=\hat{\mathbf{q}}_{ij} \cdot \mathbf{v}_{ij}$ and $F_{ij}^{C}(q_{ij})$ is the magnitude of the conservative force (\ref{eqn:Conservative_Force}). The above equation is an Ornstein-Uhlenbeck process with the exact (in the sense of distributions) solution \cite{Kloeden1992}
\begin{equation}\label{eqn:OU_exact}
    v_{ij}(t) = \frac{F_{ij}^{C}}{\tau m_{ij}} + e^{-\tau t}\left( v_{ij}(0) - \frac{F_{ij}^{C}}{\tau m_{ij}} \right) + \sqrt{ \frac{k_{B}T(1-e^{-2\tau t})}{m_{ij}} }\mathrm{R}_{ij}(t),
\end{equation}
where $\tau=\gamma \omega^{D} / m_{ij}$, $v_{ij}(0)$ are the initial relative velocities and $\mathrm{R}_{ij}(t)$ are normally distributed variables with zero mean and unit variance. Thus the increment velocities can be obtained as
\begin{equation}
    \Delta v_{ij} = v_{ij}(t) - v_{ij}(0)= \left( v_{ij}(0) - \frac{F_{ij}^{C}}{\tau m_{ij}} \right)( e^{-\tau t} - 1 ) + \sqrt{ \frac{k_{B}T(1-e^{-2\tau t})}{m_{ij}} }\mathrm{R}_{ij}(t),
\end{equation}
and the corresponding momenta can be updated by
\begin{align}
  \mathbf{p}^{n+1}_{i} &= \mathbf{p}^{n}_{i} + h m_{ij}\Delta v_{ij} \hat{\mathbf{q}}^{n}_{ij}, \\
  \mathbf{p}^{n+1}_{j} &= \mathbf{p}^{n}_{j} - h m_{ij}\Delta v_{ij} \hat{\mathbf{q}}^{n}_{ij},
\end{align}
which defines the propagator $e^{h\mathcal{L}_{\text{S}_{i,j}}}$ for each interacting pair. Overall, the propagator of the DPD-Trotter scheme can be written as
\begin{equation*}
  \exp\left(h\hat{\mathcal{L}}_\text{DPD-Trotter}\right) = \exp\left(\frac{h}{2}\mathcal{L}_\text{S}\right) \exp\left(h\mathcal{L}_\text{A}\right) \exp\left(\frac{h}{2}\mathcal{L}_\text{S}\right),
\end{equation*}
where the momentum part is defined by
\begin{equation*}
  \exp\left(\frac{h}{2}\hat{\mathcal{L}}_\text{S}\right) = \exp\left(\frac{h}{2}\mathcal{L}_{\text{S}_{N-1,N}}\right) \cdots \exp\left(\frac{h}{2}\mathcal{L}_{\text{S}_{1,3}}\right) \exp\left(\frac{h}{2}\mathcal{L}_{\text{S}_{1,2}}\right).
\end{equation*}

\subsection{Alternative Methods}
\label{Subsection:Alternative_Methods}

\subsubsection{Lowe-Andersen Thermostat}

The Schmidt number, $S_{c}$, which is the ratio of the kinematic viscosity $\nu$ to the diffusion coefficient $D$, is an important quantity that characterizes the dynamical behavior of fluids. In a typical fluid flow, water for example, momentum can be transported more rapidly than particles, resulting in Schmidt number of the order $10^{3}$. However, as it was reported in \cite{Groot1997}, the standard DPD system (\ref{eqn:DPD_system}) produces a gas-like Schmidt number of the order 1. Depending on the application, this could be a significant disadvantage for simulating more fluid-like system, although recent work by Fan \emph{et al.} \cite{Fan2006} indicates that the Schmidt number of the standard DPD system can be varied by modifying the weight function and increasing the cutoff radius.

To overcome the issue of low Schmidt number in standard DPD system, instead of using a Langevin thermostat to re-equilibrate the system, Lowe \cite{Lowe1999} employed a pairwise stochastic momentum-conserving Andersen thermostat \cite{Andersen1980}, in which after updating the positions and momenta due to the conservative forces only by using the standard velocity Verlet method, the momenta are updated, with probability $P=\Gamma h$, by reselecting the relative velocities for interacting pairs from the Maxwell-Boltzmann distribution,
\begin{align}
  \mathbf{p}_{i} &:= \mathbf{p}_{i} + \Delta \mathbf{p}_{ij}, \label{eqn:Lowe_p_n+1_i} \\
  \mathbf{p}_{j} &:= \mathbf{p}_{j} - \Delta \mathbf{p}_{ij}, \label{eqn:Lowe_p_n+1_j}
\end{align}
with
\begin{equation}\label{eqn:Lowe_delta_P_ij}
  \Delta \mathbf{p}_{ij} = m_{ij} \left( \mathrm{R}_{ij}(t)\sqrt{k_{B}T/m_{ij}} - \hat{\mathbf{q}}_{ij}\cdot \mathbf{v}_{ij} \right) \hat{\mathbf{q}}_{ij},
\end{equation}
where $\mathrm{R}_{ij}(t)$ are Gaussian random variables with zero mean and unit variance. The parameter $\Gamma$ can be thought of as the stochastic randomization frequency with upper limit $1/h$. Lowe's method is frequently referred to as the Lowe-Andersen (LA) thermostat, which still conserves the momentum and hydrodynamics. The additional Andersen thermostat does not change the distribution of the system \cite{Andersen1980}, therefore the same invariant distribution (\ref{eqn:Invariant_Distribution_DPD}) as in the standard DPD system is maintained. Most important, the LA thermostat is capable of varying the Schmidt number by changing the parameter $\Gamma$. When the probability of further updating the momentum is high (large $P$), the viscosity is high and diffusion coefficient is low, resulting in large Schmidt number in the regime of typical fluids. The LA thermostat has been applied in molecular dynamics simulation by Koopman and Lowe \cite{Koopman2006}. It is worth mentioning that the generator of the LA thermostat does not converge to that of the standard DPD system as $h \rightarrow 0$.

\subsubsection{Peters Thermostat}

Based on various numerical studies on the DPD system,  all the numerical methods based on discretization of the equation of motion are dependent on the stepsize.   In order to reduce the dependence, Peters generalized the Lowe-Andersen (LA) thermostat and presented another approach to re-equilibrate the system \cite{Peters2004}. Following the update of the conservative part by using the standard velocity Verlet scheme, the momenta of all interacting pairs will be further updated (not in random order as in the original paper, which does not have much effect on the results but reduces computational cost) as follows
\begin{align}
  \mathbf{p}_{i} &:= \mathbf{p}_{i} + \Delta \mathbf{p}_{ij}, \\
  \mathbf{p}_{j} &:= \mathbf{p}_{j} - \Delta \mathbf{p}_{ij},
\end{align}
with
\begin{equation}\label{eqn:Lowe_delta_P_ij}
  \Delta \mathbf{p}_{ij} = \left( - \gamma_{ij} (\hat{\mathbf{q}}_{ij}\cdot \mathbf{v}_{ij}) h + \sigma_{ij} \sqrt{h}\mathrm{R}_{ij}(t) \right) \hat{\mathbf{q}}_{ij},
\end{equation}
where $\mathrm{R}_{ij}(t)$ are the standard Gaussian random variables as in Lowe-Andersen thermostat, and the coefficients $\gamma_{ij}$ and $\sigma_{ij}$ satisfies the following condition
\begin{equation*}
  \sigma_{ij} = \sqrt{2k_{B}T\gamma_{ij} \left( 1 - \gamma_{ij}h/(2m_{ij}) \right) },
\end{equation*}
which reduces to the fluctuation-dissipation theorem (\ref{eqn:Fluctuation_Dissipative_Theorem}) in standard DPD in the limit of $h \rightarrow 0$. Two possible choices of the coefficients was presented in the original paper \cite{Peters2004}, but only ``Scheme II'', which has less restriction on the choice of stepsize $h$, was chosen to compare with other methods in this article. In the $h \rightarrow 0$ limit, it can be easily shown that the generator of the Peters thermostat converges to that of the standard DPD system and therefore conserves the canonical ensemble. Unfortunately, numerical simulations in the original paper showed that the thermostat still exhibits significant deviation both in static (kinetic temperature) and dynamical (diffusion coefficient) quantities in standard model settings (model B in the language of \cite{Nikunen2003}) when the timestep is above 0.05.

\subsubsection{Nos\'{e}-Hoover-Lowe-Andersen Thermostat}

Recently, Stoyanov and Groot combined the Lowe-Andersen (LA) thermostat with a Nos\'{e}-Hoover-like thermostat to construct a local Galilean invariant stochastic momentum-conserving thermostat, Nos\'{e}-Hoover-Lowe-Andersen (NHLA) thermostat \cite{Stoyanov2005}, to achieve direct kinetic temperature control. Unlike the LA thermostat, a modified version of the velocity Verlet scheme is used to update the positions and velocities at the start of each integration step
\begin{align}
  \mathbf{q}_{i} &:= \mathbf{q}_{i} + h\mathbf{v}_{i} + h^{2}\mathbf{F}_{i}^{C}(\mathbf{q})/2, \\
  \mathbf{v}_{i} &:=  \mathbf{v}_{i} + h\mathbf{F}_{i}^{C}(\mathbf{q})/2,
\end{align}
which is followed by the calculation of the updated conservative forces. After that, the fraction $(1-P)$ of interacting pairs that do not have their relative velocities stochastically reselected are thermalized by a deterministic method instead. For each such pair, the dissipative force is calculated
\begin{equation*}
   \mathbf{F}_{ij}^{D} = \alpha \omega^{R}(q_{ij}) (\hat{\mathbf{q}}_{ij}\cdot \mathbf{v}_{ij})\hat{\mathbf{q}}_{ij},
\end{equation*}
where $\alpha$ is a coupling parameter chosen as $0.9/(\rho h)$ in this article such that, overall, the dissipative force defined above is the same as that in the original paper \cite{Stoyanov2005}, and $\rho$ is the particle density. The dissipative force of each particle is updated
\begin{align}
  \mathbf{F}_{i}^{D} &:= \mathbf{F}_{i}^{D} + \mathbf{F}_{ij}^{D}, \\
  \mathbf{F}_{j}^{D} &:= \mathbf{F}_{j}^{D} - \mathbf{F}_{ij}^{D}.
\end{align}
Then, after the further update of the velocities
\begin{equation}
   \mathbf{v}_{i} :=  \mathbf{v}_{i} + h\mathbf{F}_{i}^{C}(\mathbf{q})/2,
\end{equation}
the momenta are corrected by
\begin{equation}\label{eqn:NHLA_P_update}
  \mathbf{p}_{i} := \mathbf{p}_{i} + h(1-\tilde{T}_{k}/T_{0})\mathbf{F}_{i}^{D},
\end{equation}
where $T_{0}$ is the desired temperature and the momentary kinetic temperature $\tilde{T}_{k}$ is calculated from the relative velocities at the time of calculating the conservative force to enhance computational efficiency and save memory space (this is slightly different from the approach in the original paper which uses the updated Verlet neighbor lists but the stored velocities from the previous integration step)
\begin{equation}\label{eqn:momentary_KT}
  k_{B}\tilde{T}_{k} = \frac{\sum_{j\neq i}\omega(q_{ij})m_{ij}(\mathbf{v}_{i}-\mathbf{v}_{j})^{2}}{3\sum_{j\neq i}\omega(q_{ij})},
\end{equation}
where $\omega(q_{ij})$ is a smearing function chosen as
\begin{equation}\label{eqn:NHLA_smearing_function}
  \omega(q_{ij})=
  \begin{cases}
  1, & q_{ij}<r_{c};\\
  0, & q_{ij}\geq r_{c}.
  \end{cases}
\end{equation}
Finally, the momenta are further updated as in the LA thermostat (Eqs. (\ref{eqn:Lowe_p_n+1_i})-(\ref{eqn:Lowe_p_n+1_j})).

The factor $(1-\tilde{T}_{k}/T_{0})$ in (\ref{eqn:NHLA_P_update}) acts like the ``friction coefficient'' to tune the kinetic temperature of the system. It is actually not a dynamical variable as in the standard Nos\'{e}-Hoover thermostat, instead is more closely related to the Berendsen thermostat \cite{Berendsen1984}. As reported in \cite{Stoyanov2005}, the NHLA thermostat maintains an order of magnitude improvement in kinetic temperature and can also vary the Schmidt number by several orders of magnitude as in the LA thermostat. However, with large stepsizes that maintain good kinetic temperature control (1\% relative error) in the NHLA thermostat, substantial errors in configurational temperature were reported \cite{Allen2006}, which indicates that the system temperature was not sampled correctly. It is worthy of mention that a slightly modified integration strategy was used in \cite{Allen2006}, which does not have much effect on the results as discussed in the original paper \cite{Stoyanov2005}. Moreover, the generator of the NHLA thermostat does not converge to that of the standard DPD system as $h \rightarrow 0$, and, it has not been shown that the NHLA thermostat preserves the canonical ensemble.

\section{Extended Variable Momentum-Conserving Thermostats}
\label{Section:Extended_Variable}

\subsection{Pairwise Nos\'{e}-Hoover Thermostat}

In all the standard DPD methods (Section \ref{Subsection:Numerical_Integration_Schemes}) and alternative methods (Section \ref{Subsection:Alternative_Methods}) in DPD simulations, a random number has to be generated for each interacting pair, which can be very time-consuming depending on the particle density and cutoff radius, and, becomes trickier when parallel computing techniques (multiple processors for domain-decomposed cells) are used: it requires additional, even substantial, effort to communicate interacting particles in different cells \cite{Phillips2011,Afshar2013}. Based on the Nos\'{e}-Hoover-Lowe-Andersen (NHLA) thermostat by Stoyanov and Groot \cite{Stoyanov2005}, Allen and Schmid  \cite{Allen2007} presented a new thermostat of the Nos\'{e}-Hoover type, in which stochastic terms were totally removed and the constant friction coefficient was replaced by a dynamical variable that was driven by the difference between the instantaneous kinetic temperature and the target temperature. The so-called pairwise Nos\'{e}-Hoover (PNH) thermostat conserves the momentum and is also Galilean-invariant, therefore correct hydrodynamics are still expected to be generated and it can be used in DPD simulations. Moreover, one may find it useful in nonequilibrium molecular dynamics (NEMD) to reduce unphysical behaviors (see more discussions in \cite{Allen2007}).

The equation of motion of the PNH thermostat (for particle $i$) is given by
\begin{equation}
  \label{eqn:PNH_system}
  \begin{aligned}
    \dd \mathbf{q}_{i} &= m_{i}^{-1}\mathbf{p}_{i}\dd t, \\
    \dd \mathbf{p}_{i} &= \mathbf{F}^{C}_{i}(\mathbf{q})\dd t - \xi \mathbf{V}_{i}(\mathbf{q},\mathbf{p})\dd t, \\
    \dd \xi            &= G(\mathbf{q},\mathbf{p})\dd t,
  \end{aligned}
\end{equation}
where $\xi$ is the dynamical variable and $G(\mathbf{q},\mathbf{p})$ is the instantaneous accumulated deviation of the kinetic temperature away from the target temperature \cite{Allen2007}
\begin{equation}
  \label{eqn:PNH_system_G}
    G(\mathbf{q},\mathbf{p}) = {\mu}^{-1}\sum_{i}\sum_{j>i}\omega^{D}(q_{ij}) \left[ \left( \mathbf{v}_{ij}\cdot\hat{\mathbf{q}}_{ij} \right)^{2} - k_{B}T/m_{ij}  \right],
\end{equation}
where $\mu$ is a coupling parameter which is referred to as the ``thermal mass''. Canonical ensemble is still preserved with modified invariant distribution than $\rho_{\beta}$ (\ref{eqn:Invariant_Distribution_DPD}) in standard DPD system
\begin{equation}\label{eqn:Invariant Distribution PNH}
  \hat{\rho}_{\beta}(\mathbf{q},\mathbf{p},\xi) = \frac{1}{\hat{Z}}\exp\left({-\beta H(\mathbf{q},\mathbf{p})}\right)\exp\left({-\beta\mu\xi^{2}/2}\right).
\end{equation}

A nonsymmetric integration algorithm (see Appendix \ref{Section:Appendix_Schemes}) was applied in the original paper \cite{Allen2007} to solve the system.

\subsection{Pairwise Nos\'{e}-Hoover-Langevin Thermostat}

In order to enhance the ergodicity and have a better temperature control, we have reformulated the pairwise Nos\'{e}-Hoover (PNH) thermostat to form a pairwise Nos\'{e}-Hoover-Langevin (PNHL) thermostat by adding a Langevin thermostat to the additional variable $\xi$ in such a way that the invariant distribution (\ref{eqn:Invariant Distribution PNH}) is not violated. As in the PNH thermostat, the PNHL thermostat has the potential of being useful in NEMD, but we focus on the application of DPD in this article.

The SDEs of the PNHL system (for particle $i$) is given by
\begin{equation}
  \label{eqn:PNHL_system}
  \begin{aligned}
    \dd \mathbf{q}_{i} &= m_{i}^{-1}\mathbf{p}_{i}\dd t, \\
    \dd \mathbf{p}_{i} &= \mathbf{F}^{C}_{i}(\mathbf{q})\dd t - \xi \mathbf{V}_{i}(\mathbf{q},\mathbf{p})\dd t, \\
    \dd \xi            &= G(\mathbf{q},\mathbf{p})\dd t - \gamma^{\ast} \xi \dd t + \sigma^{\ast} \dd \mathrm{W},
  \end{aligned}
\end{equation}
where coefficient constants $\gamma^{\ast}$ and $\sigma^{\ast}$ satisfy the fluctuation-dissipation theorem in standard Langevin dynamics
\begin{equation}
  {\sigma^{\ast}}^{2}=2\gamma^{\ast} k_{B}T/\mu,
\end{equation}
and $\mathrm{W}=\mathrm{W}(t)$ is a standard Wiener process.

The vector field of the PNHL system can be split into five pieces below such that each piece can be solved ``exactly'',
\begin{equation}
  \dd \left[ \begin{array}{c} \mathbf{q}_{i} \\ \mathbf{p}_{i} \\ \xi \end{array} \right] = \underbrace{\left[ \begin{array}{c} m_{i}^{-1}\mathbf{p}_{i} \\ \mathbf{0} \\ 0 \end{array} \right] \dd t}_\text{A} + \underbrace{\left[ \begin{array}{c} \mathbf{0} \\ \mathbf{F}^{C}_{i} \\ 0 \end{array} \right] \dd t}_\text{B} + \underbrace{\left[ \begin{array}{c} \mathbf{0} \\ - \xi \mathbf{V}_{i} \\ 0 \end{array} \right] \dd t}_\text{C} + \underbrace{\left[ \begin{array}{c} \mathbf{0} \\ \mathbf{0} \\ G \end{array} \right] \dd t}_\text{D} + \underbrace{\left[ \begin{array}{c} \mathbf{0} \\ \mathbf{0} \\ - \gamma^{\ast} \xi \dd t + \sigma^{\ast} \dd \mathrm{W} \end{array} \right]}_\text{O}.
\end{equation}
Note that the operators of A and B are exactly the same as defined in (\ref{eqn:generator_A}) and (\ref{eqn:generator_B}), respectively. We can also write down the operators for the remaining pieces as
\begin{equation*}\label{eqn:operators_C_D_O}
  \begin{aligned}
    \mathcal{L}_\text{C} &= -\xi \sum_{i} \mathbf{V}_{i}(\mathbf{q},\mathbf{p}) \cdot \nabla_{\mathbf{p}_{i}}, \\
    \mathcal{L}_\text{D} &= G(\mathbf{q},\mathbf{p}) \frac{\partial}{\partial\xi}, \\
    \mathcal{L}_\text{O} &= - \gamma^{\ast}\xi\frac{\partial}{\partial\xi} + \frac{{\sigma^{\ast}}^{2}}{2}\frac{\partial^{2}}{\partial\xi^{2}}.
  \end{aligned}
\end{equation*}
The generator O here is slightly different from that in (\ref{eqn:generator_DPD_O}) which involves pairwise terms in DPD.  Overall, the generator for the PNHL system can be written as
\begin{equation}\label{eqn:operator_PNHL_dagger}
  \mathcal{L}_\text{PNHL} = \mathcal{L}_\text{A} + \mathcal{L}_\text{B} + \mathcal{L}_\text{C} + \mathcal{L}_\text{D} + \mathcal{L}_\text{O}.
\end{equation}

There are a variety of approaches to splitting this system.   For example, we could use the same technique as in DPD-Trotter scheme (Section \ref{Subsubsection:DPD_Trotter}) to solve part C, but without the conservative and stochastic terms. Also, the O part may be solved exactly using the analytical weak solution of the Ornstein-Uhlenbeck process \cite{Kloeden1992}
\begin{equation}
    \xi(t) = e^{-\gamma^{\ast} t}\xi(0) + \sqrt{ k_{B}T(1-e^{-2\gamma^{\ast} t})/\mu }\mathrm{R}(t),
\end{equation}
where $\xi(0)$ is the initial value of the additional variable and $\mathrm{R}(t)$ are uncorrelated independent standard normal random variables.

Interestingly as noted in the setting of Langevin dynamics \cite{Leimkuhler2013,Leimkuhler2013a}, integrating those different splitting pieces in different orders gives dramatically different performances in terms of kinetic temperature control and other configurational quantities. We present here two approaches to integrate the PNHL system: first in a symmetric manner, termed PNHL-S, and the other nonsymmetric, termed PNHL-N. The propagators of the two schemes (see more details in Appendix \ref{Section:Appendix_Schemes}) may be defined as
\begin{equation*}
  e^{h\hat{\mathcal{L}}_\text{PNHL-S}} = e^{\frac{h}{2}\mathcal{L}_\text{A}} e^{\frac{h}{2}\mathcal{L}_\text{B}} e^{\frac{h}{2}\mathcal{L}_\text{C}} e^{\frac{h}{2}\mathcal{L}_\text{D}} e^{h\mathcal{L}_\text{O}} e^{\frac{h}{2}\mathcal{L}_\text{D}} e^{\frac{h}{2}\mathcal{L}_\text{C}} e^{\frac{h}{2}\mathcal{L}_\text{B}} e^{\frac{h}{2}\mathcal{L}_\text{A}},
\end{equation*}
and
\begin{equation*}
  e^{h\hat{\mathcal{L}}_\text{PNHL-N}} = e^{\frac{h}{2}\mathcal{L}_\text{A}} e^{\frac{h}{2}\mathcal{L}_\text{B}} e^{\frac{h}{2}\mathcal{L}_\text{C}} e^{\frac{h}{2}\mathcal{L}_\text{D}} e^{h\mathcal{L}_\text{O}} e^{\frac{h}{2}\mathcal{L}_\text{D}} e^{\frac{h}{2}\mathcal{L}_\text{C}} e^{\frac{h}{2}\mathcal{L}_\text{A}} e^{\frac{h}{2}\mathcal{L}_\text{B}}.
\end{equation*}
It is important to mention that the only difference between these two integrators is the order of integrating the last two pieces. In particular, an additional force calculation is needed in the PNHL-N scheme just before updating the last  B piece at the end of each integration step.   In experiments, the high per-timestep cost of PNHL-N was found to be offset by a great increase in accuracy and usable steplength.   Detailed numerical comparisons will be presented in Section \ref{Section:Numerical_Experiments}.

\section{Error in DPD Simulations}
\label{Section:Error_in_DPD}

To our knowledge, little attention has been paid to the mathematical analysis of the DPD system, or more generally stochastic momentum-conserving thermostats. Since the spectral properties of the Fokker-Planck operators in the case of DPD is not available, a rigorous study of the order of convergence of numerical methods in this context has been lacking. Because of the inclusion of stochastic terms, it is not reliable to depend directly on intuition regarding the error behavior of deterministic schemes \cite{Leimkuhler2005,Hairer2006}.  Here we perform a few first steps  toward the analysis of stochastic DPD integrators by extending a framework for investigating the perturbation of long-time average computed using numerical methods in Langevin dynamics proposed recently by Leimkuhler and Matthews \cite{Leimkuhler2013,Leimkuhler2013a}.

In DPD simulations, the error in averages related to the evolving distribution is generally of interest, i.e. the {\em weak error} (finite-time error in the weak sense) for nonequilibrium and dynamical properties, and, {\em long-time error} (error in the invariant distribution obtained as $t\rightarrow \infty$) for thermodynamics.  We next give definitions of these two errors following \cite{Leimkuhler2014}.

For the weak error, consider a finite time interval $[0,\tau]$ with $\tau=nh$. The probability measure associated with certain system is described by a probability density $\rho(\mathbf{z},t)$ which evolves according to the Fokker-Planck equation
\begin{equation*}
  \frac{ \partial \rho}{ \partial t} = \mathcal{L}^{\dag}\rho,
\end{equation*}
where $\mathcal{L}^{\dag}$ is the adjoint of the generator of the system (the Fokker-Planck operator of the standard DPD system is given in \cite{Espanol1995}). Assuming ergodicity, the solution $\rho(\mathbf{z},t)$ evolves from an initial probability distribution $\rho(\mathbf{z},0)$ to the steady state (invariant distribution) $\rho(\mathbf{z},\infty)=\rho_\beta$. For a smooth and bounded test function $\phi$ of a suitable class, the average of $\phi$ at time $\tau$ may be defined  by
\begin{equation}\label{eqn:average_exact_finite_time}
  \bar{\phi}(\tau) = \int \phi(\mathbf{z}) \rho(\mathbf{z},\tau) \dd \mathbf{z}.
\end{equation}
The discretization scheme can also be viewed as giving rise to an evolving sequence of probability distributions $\rho_{1},\rho_{2},\cdots$. With stepsize $h$, the average at time $\tau=nh$ is defined as
\begin{equation}\label{eqn:average_approx_finite_time}
  \hat{\phi}(\tau,h) = \int \phi(\mathbf{z}) \rho_{n}(\mathbf{z}) \dd \mathbf{z}.
\end{equation}
Thus, we could define the \emph{weak error} as the difference between (\ref{eqn:average_exact_finite_time}) and (\ref{eqn:average_approx_finite_time})
\begin{equation}\label{eqn:weak_error}
  | \bar{\phi}(\tau) - \hat{\phi}(\tau,h) | \leq K(\tau)h^{p},
\end{equation}
where the coefficient  $K$ depends on the time interval and $p$ is the order of the weak error. To be more precise, $K$ also depends on the initial distribution $\rho(\mathbf{z},0)$ as well as the particular observable $\phi$. The asymptotic ($\tau \rightarrow \infty$) behavior \cite{Talay1990} of $K$ can be used to describe the performance of the numerical method for computing averages with respect to the invariant distribution. Hence, the \emph{long-time error} in averages can be written as
\begin{equation}\label{eqn:long_time_error}
  \lim_{\tau \rightarrow \infty} | \bar{\phi}(\tau) - \hat{\phi}(\tau,h) | \leq K(\tau)h^{p}.
\end{equation}

\begin{table}[htbp]
\begin{center}
\resizebox{1.0\textwidth}{!}{\begin{minipage}{\textwidth}
    \begin{tabular}{ |c|c|c|c|c|c|c|c|c|c| }
    \hline
    \textbf{\footnotesize Method} & \footnotesize DPD-VV & \footnotesize DPD-S1 & \footnotesize DPD-Trotter & \footnotesize LA  & \footnotesize Peters & \footnotesize NHLA & \footnotesize PNH & \footnotesize PNHL-S & \footnotesize PNHL-N \\ \hline
    \textbf{\footnotesize Order}  & \small $\geq 1(2)$ & \small 2 & \small 2 & \small 2 & \small 2 & \small $\geq 1(2)$ & \small $\geq 1(2)$ & \small 2 & \small $\geq 1(2)$ \\
    \hline
    \end{tabular}
\caption[Table caption text]{Orders of accuracy of the \emph{long-time error} for kinetic and configurational temperatures, and, average potential energy in various methods are summarized. Theoretical values based on properties of the discretization scheme have been verified using numerical experiments. Where the theoretically expected result differs from the numerical result, we give the numerically observed convergence order in parentheses.}
\label{table:order}
\end{minipage} }
\end{center}
\end{table}

According to the Baker-Campbell-Hausdorff (BCH) formula, a nonsymmetric splitting method generally gives only first order convergence for the invariant measure (\emph{long-time error}), whereas second order is anticipated in symmetric splittings in the asymptotic limit of small stepsize \cite{Leimkuhler2013c}.  In our numerical studies, we have verified second order convergence for a number of nonsymmetric methods (Shardlow's splitting method, Lowe-Andersen thermostat and Peters thermostat) which are similar to the family of Geometric Langevin Algorithms (GLA) following \cite{Bou-Rabee2010} (see more details in Appendix \ref{Section:Appendix_Order}). We compute the \emph{long-time error} of various observables, including kinetic and configurational temperatures and average potential energy, and demonstrate the convergence order for each method in Table \ref{table:order}. The results shown in Table \ref{table:order} are based on the numerical experiments in Section \ref{Section:Numerical_Experiments}.  All the symmetric methods show second order as well as those three nonsymmetric ones (DPD-S1, LA and Peters).   Some other nonsymmetric methods, which are not of GLA type,  exhibit second order convergence in calculated quantities; this observation remains to be demonstrated rigorously.  It is entirely possible that the superconvergence observed in these special cases is related to the form of the observable we have used in our simulation test.

\section{Numerical Experiments}
\label{Section:Numerical_Experiments}

To investigate the performance of all the numerical methods described in this article, we perform systematic numerical experiments in this section.

\subsection{Simulation Details}
\label{Subsection:Simulation_Details}

Tests have been carried out by using the standard parameter set \cite{Groot1997} that is commonly used in algorithms tests \cite{Vattulainen2002,Nikunen2003,Shardlow2003,Peters2004,Stoyanov2005,Serrano2006}. A system of $N=500$ identical particles ($m_{i}=m=1$) was simulated in a cubic box (length $L=5$) with periodic boundary conditions (particle density $\rho=4$). The cut-off radius is $r_{c}=1$ and $k_{B}T=1$. The potential repulsion parameter $a_{ij}$ was set to 25, while dissipative strength $\gamma$ was chosen as 4.5, resulting the value of random strength $\sigma$ to be 3. It is worthy of mention that we did investigate the influence of other values of the friction coefficient, such as $\gamma=0.5$ and $\gamma=40.5$, but not all the results will be presented unless necessary. The initial positions of the particles were IID (independent, identically distributed) uniformly distributed over the box, while the initial momenta were IID normally distributed with mean zero and variance $k_{B}T$. Verlet neighbor lists \cite{Verlet1967} were used in all the simulations as long as possible.

In particular, the thermal mass $\mu$ in PNH and PNHL thermostats has to be chosen with care. For PNH thermostat, we used $\mu=200$ to maintain as good stability as possible for the integration scheme, while $\mu=10$ and $\gamma^{\ast}=4.5$ were used in PNHL thermostat. Since the focus of this article is on DPD simulations, the stochastic randomization frequency $\Gamma$ in LA and NHLA thermostats was set to be 0.44 as in \cite{Nikunen2003,Jakobsen2005} so that similar translational diffusion properties of the fluid were obtained.

\subsection{Measured Physical Quantities}
\label{Subsection:Physical_Quantities}

\subsubsection{Static}

The kinetic temperature $T_{k}$ appears to be the most popular quantity to be tested in DPD literature, which is defined as
\begin{equation*}
  k_{B}T_{k} = \frac{1}{d(N-1)}\sum_{i}\frac{\mathbf{p}^{2}_{i}}{m_{i}}.
\end{equation*}
But in practice, the kinetic temperature is not that important, instead those configurational quantities play more crucial roles. Recent studies \cite{Rugh1997,Butler1998} in simulations showed that the system temperature can be measured from static snapshots of its constituents' instantaneous configurations rather than their momenta. We test the configurational temperature $T_{c}$ \cite{Hirschfelder1960}, which can be defined as
\begin{equation*}
  k_{B}T_{c} = \frac{\sum_{i}\langle {\| \nabla_{i}U \|}^{2} \rangle}{\sum_{i}\langle \nabla^{2}_{i}U \rangle},
\end{equation*}
where the angle brackets denote the averages, $\nabla_{i}U$ and $\nabla^{2}_{i}U$ are respectively the gradient and Laplacian of the potential energy $U$ with respect to the position of particle $i$. To test the correctness of codes and/or algorithms in simulations, both kinetic and configurational temperatures can be used to calculate the system temperature (in principle they should both be equal to the target). However, it turned out that the configurational temperature is more reliable  \cite{Rugh1997,Butler1998} since it can rapidly and accurately track changes in system temperature even when the system is not in global thermodynamic equilibrium. It becomes more crucial when it comes to experimental studies of soft condensed matter systems \cite{Han2004,Han2005} most notably due to their applicability to overdamped systems whose instantaneous momenta may not be accessible. It was den Otter and Clarke that first investigated the deviations of the kinetic and configurational temperatures from the system temperature in DPD system \cite{DenOtter2001}. Since then, little attention has been paid to the configurational temperature in DPD simulations until Allen recently argued that the configurational temperature should be added to the list of diagnostic tests applied to DPD simulations \cite{Allen2006}. In addition, we also calculate the average potential energy $\langle U \rangle$ and the radial distribution function (RDF) $g(r)$ \cite{Allen1989,Frenkel2001}, both of which are very important configurational quantities in simulations.

\begin{figure}[tb]
\centering
\includegraphics[scale=0.5]{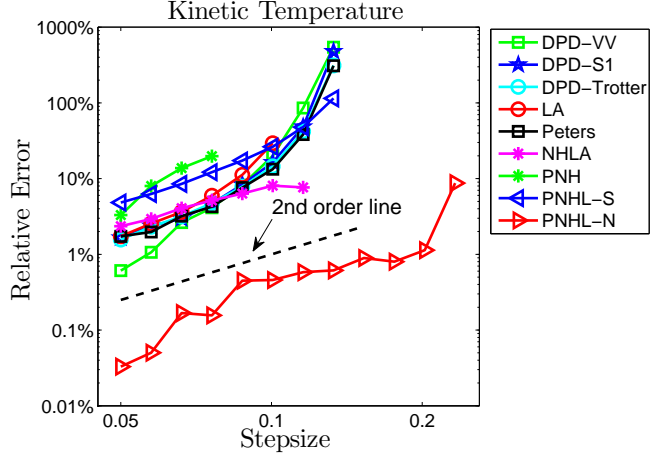}
\caption{\small (Color.) Log-log plot of the relative error in kinetic temperature against stepsize by using various numerical methods. The system was simulated for 1000 reduced time units but only last 80\% data were collected to calculate the static quantity to make sure the system was well equilibrated. The stepsizes tested began at $h=0.05$ and were increased incrementally by 15\% until all methods either were above 100\% relative error or became unstable.}
\label{fig:Comp_KT}
\end{figure}

\subsubsection{Dynamical}
\label{Subsubsection:Dynamical}

To have a deeper understanding of how the physical system evolves, it is not enough to just evaluate the static quantities described above. It is essential to measure and compare the relevant dynamical properties.  In general, in simulation, one relies on various Green-Kubo formulas \cite{Evans2008} to calculate various transport coefficients; in this article we restrict ourselves to two special cases.

The velocity autocorrelation function (VAF) is an important measure of dynamical fidelity that numerical methods should be able to reproduce, particularly if they are to be used for nonequilibrium (transient-phase simulation). The VAF, which characterizes the translational diffusion of the system, is defined as
\begin{equation*}
  \psi(t) = \langle \mathbf{v}_{i}(t) \cdot \mathbf{v}_{i}(0) \rangle,
\end{equation*}
where $\mathbf{v}_{i}(0)$ is the initial velocity picked up after the system is well equilibrated. By integrating the VAF (Green-Kubo relation \cite{Green1954,Kubo1957}), we can compute the translational diffusion coefficient, which can also be obtained by using Einstein's relation \cite{Einstein1905} giving the diffusion coefficient as the slope of the mean square displacement as a function of time $t$.

To investigate the rotational relaxation process of the system, we measure the transverse momentum autocorrelation function (TMAF) \cite{Tang1995,Hansen2006}, which is related to the shear viscosity, $\eta$, in the hydrodynamic limit, i.e. small wavenumber $k$ and large time $t$, and is defined by
\begin{equation*}
  C(k,t) = \langle p_{x}(k,t) p_{x}(-k,0) \rangle \propto \exp \left(-\frac{k^{2}\eta}{\rho m}t \right),
\end{equation*}
where $\rho$ is the particle density, $m$ is the particle mass and $p_{x}(k,t)$ represents the transverse component of the momentum,
\begin{equation*}
  p_{x}(k,t)= \sum^{N}_{j=1}p_{j,x}(t) \exp(i k q_{j,z}(t)),
\end{equation*}
where $p_{j,x}$ denotes the $x$-component of the momentum of particle $j$ and similarly $q_{j,z}$ represents the $z$-component of its position (see more details in \cite{Tang1995}). Note that, $i$ here is the imaginary unit in complex number.

\begin{figure}[tb]
\centering
\includegraphics[scale=0.5]{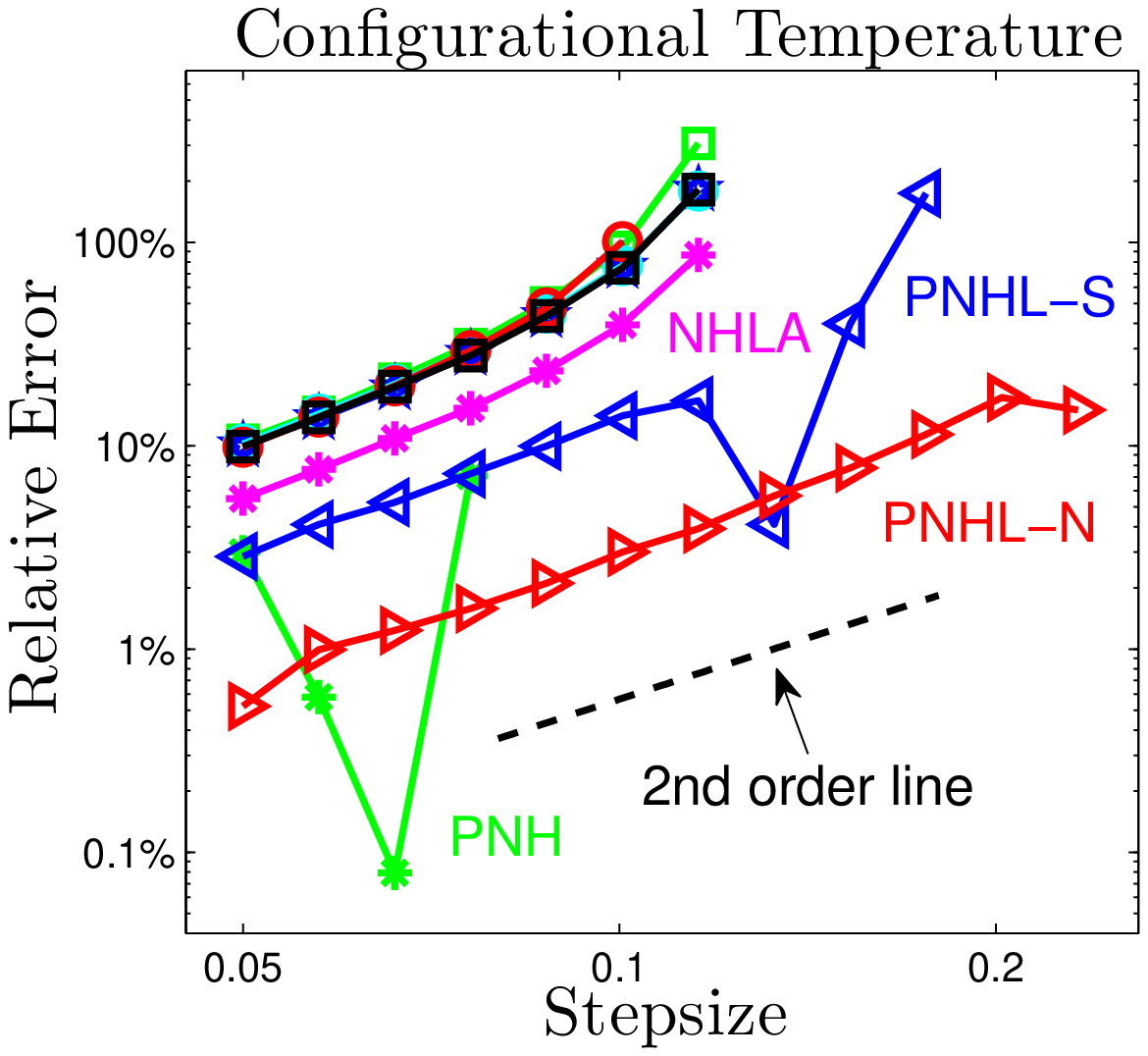}
\includegraphics[scale=0.5]{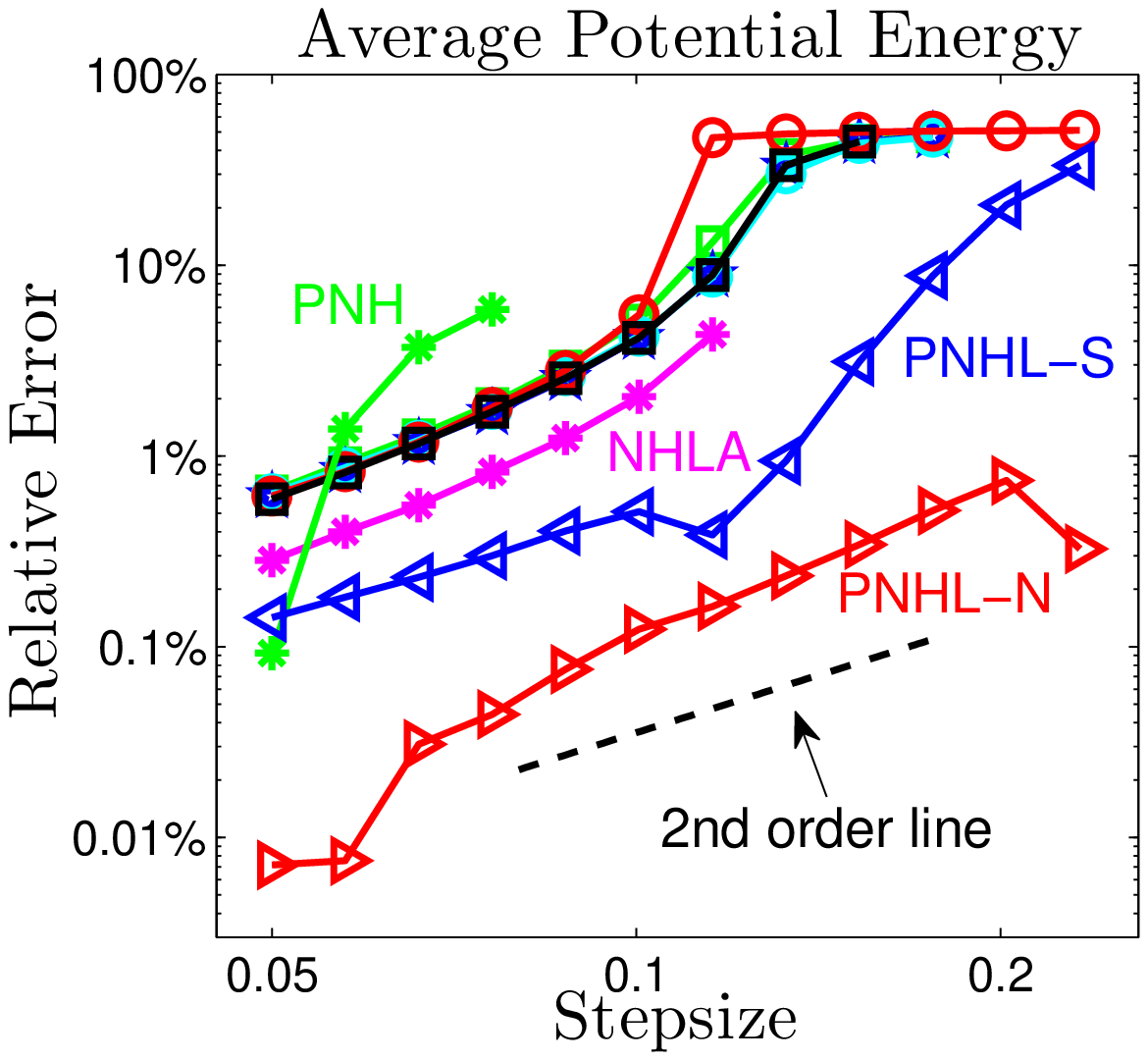}
\caption{\small (Color.) Comparisons of the relative error in configurational temperature (left) and average potential energy (right) against stepsize by using various numerical methods. The format of the plots is the same as in Fig. \ref{fig:Comp_KT}. Most of the methods show similar behaviors except NHLA (magenta asterisks), PNH (green asterisks), PNHL-S (blue triangles) and PNHL-N (red triangles).}
\label{fig:Comp_CT_U}
\end{figure}

\subsection{Results}
\label{Subsection:Results}

The kinetic temperature control of various methods was tested and shown in Fig. \ref{fig:Comp_KT}. According to the black dashed second order line in the figure, all the methods seem to have second order convergence to the invariant measure, which verifies the error analysis results on nonsymmetric DPD-S1, LA and Peters thermostats in Section \ref{Section:Error_in_DPD}, but is a bit surprising for DPD-VV, NHLA, PNH and PNHL-N methods that were also based on nonsymmetric splittings.

The performance of standard DPD methods (DPD-VV, DPD-S1 and DPD-Trotter) and the Peters thermostat that converges to the standard DPD system in the limit of $h\rightarrow0$, are almost indistinguishable with the tendency to quickly blow up after the stepsize reaches $h=0.1$. Both LA and NHLA thermostats show similar behaviors and maximal stepsizes that can be used are limited around $h=0.1$. The latter displays a better accuracy when stepsize is large due to the additional Nos\'{e}-Hoover-like thermostat (\ref{eqn:NHLA_P_update}). The PNH thermostat illustrates the worst kinetic temperature accuracy and we can also see the clear low stability threshold for the PNH thermostat due to the lack of ergodicity. Most surprisingly and interestingly, the PNHL-S and PNHL-N methods, based on different splitting orders on the same pairwise Nos\'{e}-Hoover-Langevin (PNHL) thermostat (\ref{eqn:PNHL_system}), show dramatically different kinetic temperature control: the symmetric PNHL-S method is worse than most of the methods, whereas the nonsymmetric PNHL-N method outstandingly maintains almost two orders of magnitude improvement on the accuracy of kinetic temperature than all the other methods.

\begin{figure}[tb]
\centering
\includegraphics[scale=0.5]{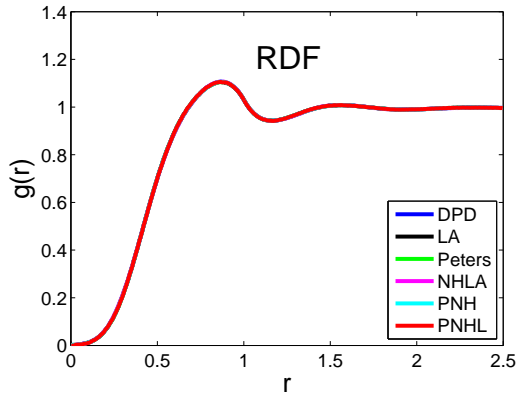}
\caption{\small (Color.) Comparisons of radial distribution function (RDF) $g(r)$ of various numerical methods by using very small stepsize $h=0.01$. All the methods exhibit exactly the same RDF. }
\label{fig:Comp_RDF}
\end{figure}

Configurational quantities (configurational temperature and average potential energy) were compared in Fig. \ref{fig:Comp_CT_U}. Again, all the methods seem to show second order convergence to the invariant measure except the PNH thermostat, which displays a stability threshold of $h=0.05$. Most of the methods, including standard DPD methods, LA and Peters thermostats, are indistinguishable and cross the 100\% barrier in configurational temperature and 10\% barrier in average potential energy respectively around $h=0.1$. As in the case of kinetic temperature, the performance of the NHLA thermostat on those configurational quantities is slightly better than the majority due to the additional thermostat. Unlike the kinetic temperature case, the PNHL-S method does have very good accuracy in configurational quantities with almost one order of magnitude improvement than the majority. Incredibly, the PNHL-N method manages more than one order of magnitude accuracy enhancement in configurational temperature and almost two orders of magnitude in average potential energy.

\begin{figure}[htb]
\centering
\includegraphics[scale=0.3]{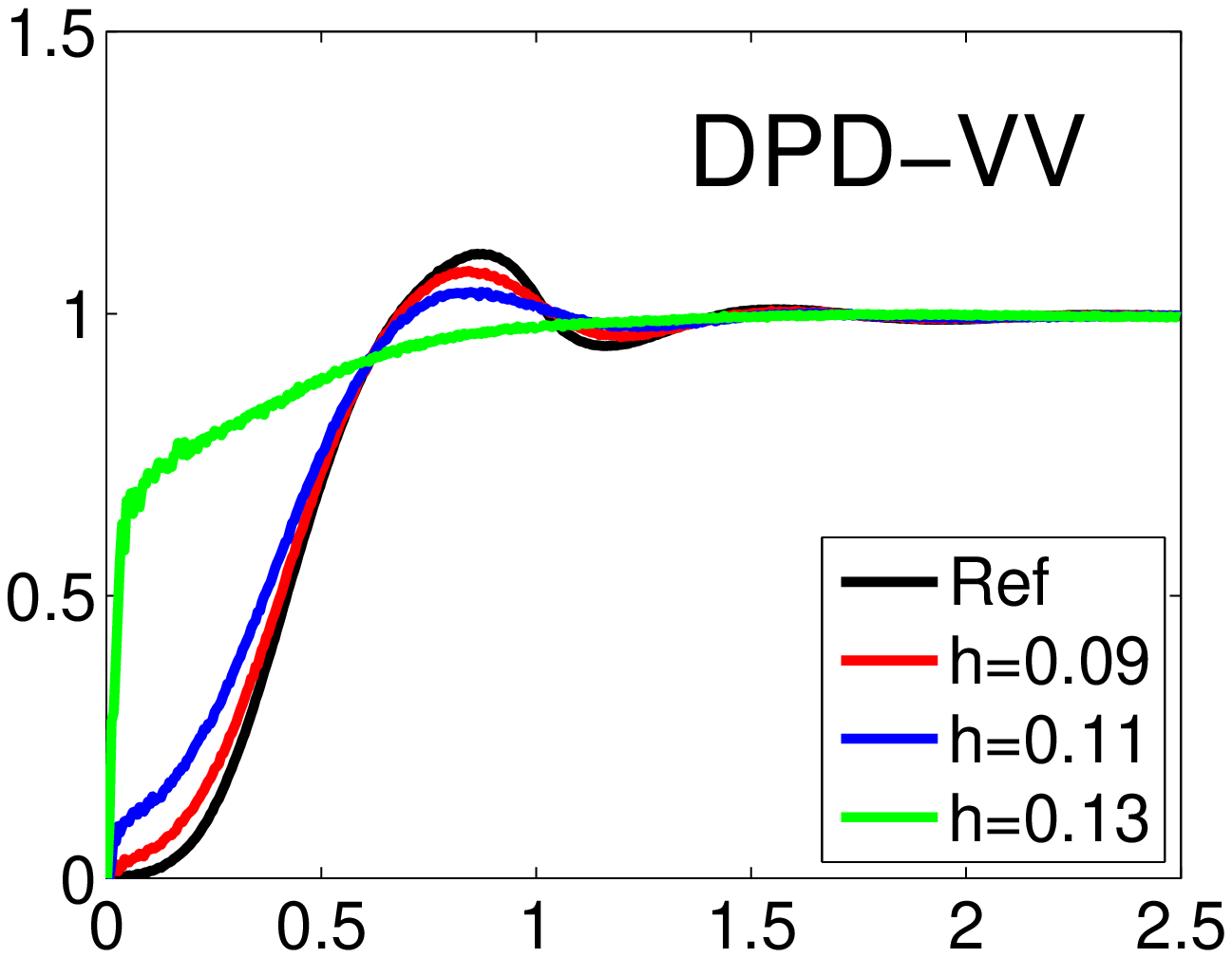}
\includegraphics[scale=0.3]{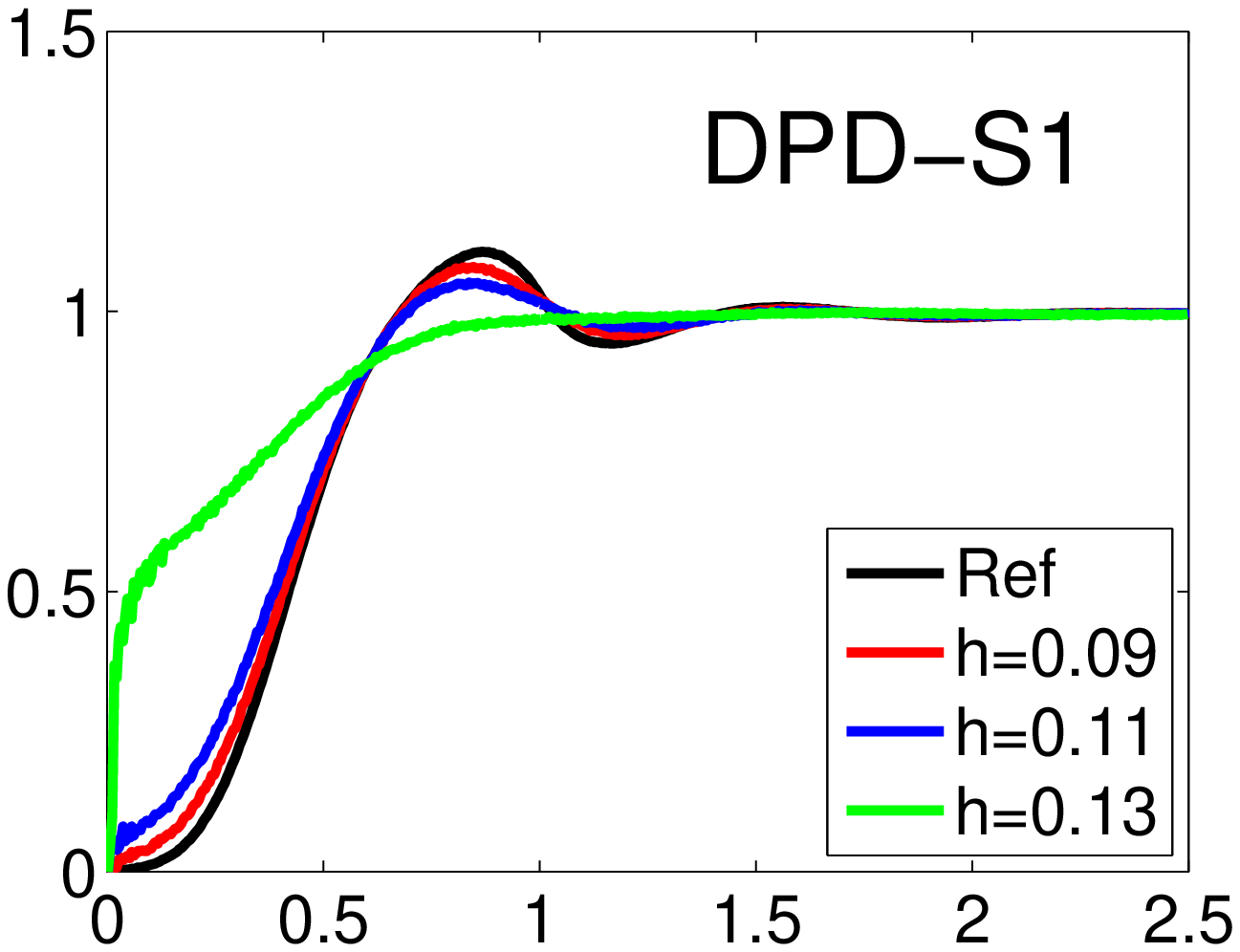}
\includegraphics[scale=0.3]{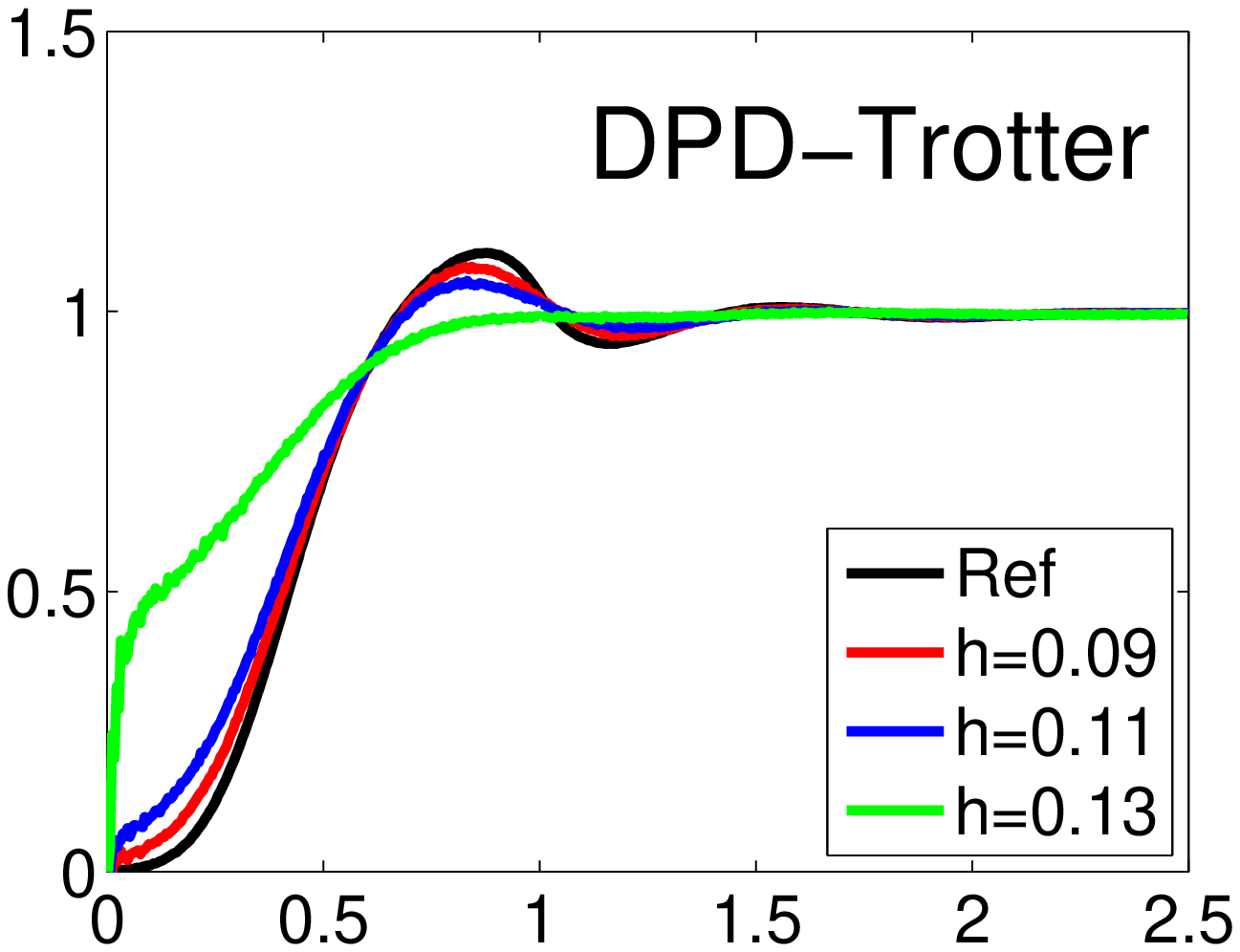}
\includegraphics[scale=0.3]{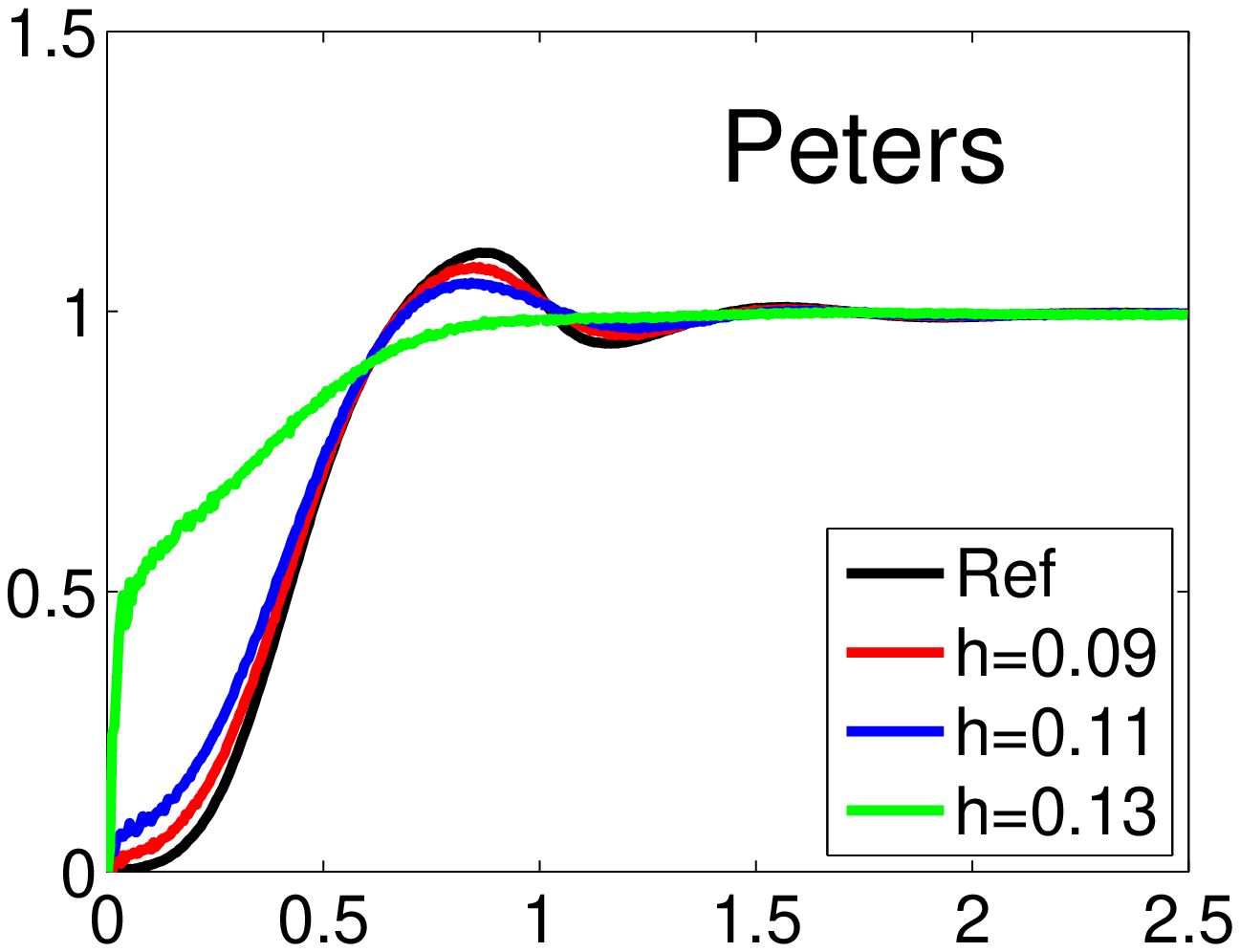}
\includegraphics[scale=0.3]{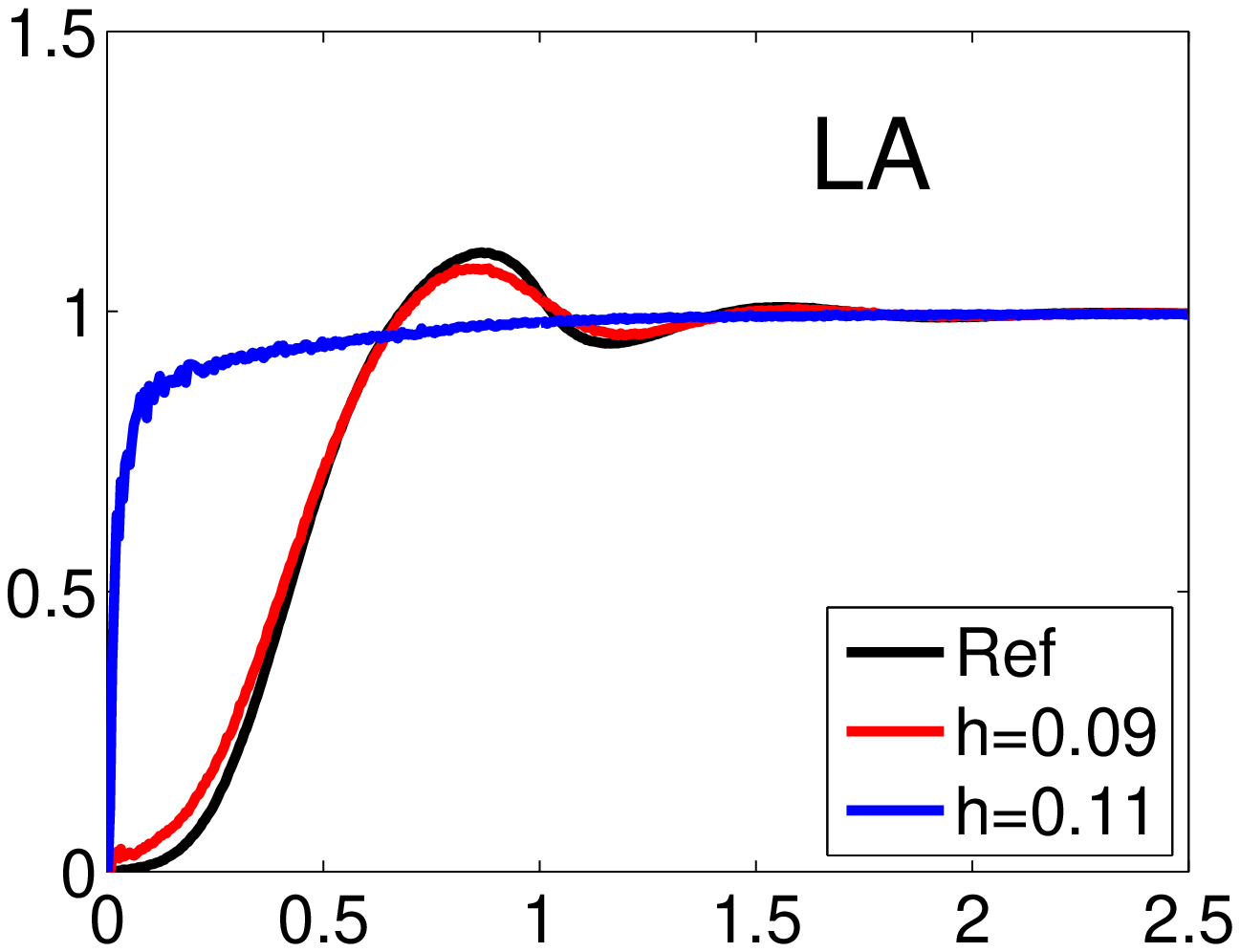}
\includegraphics[scale=0.3]{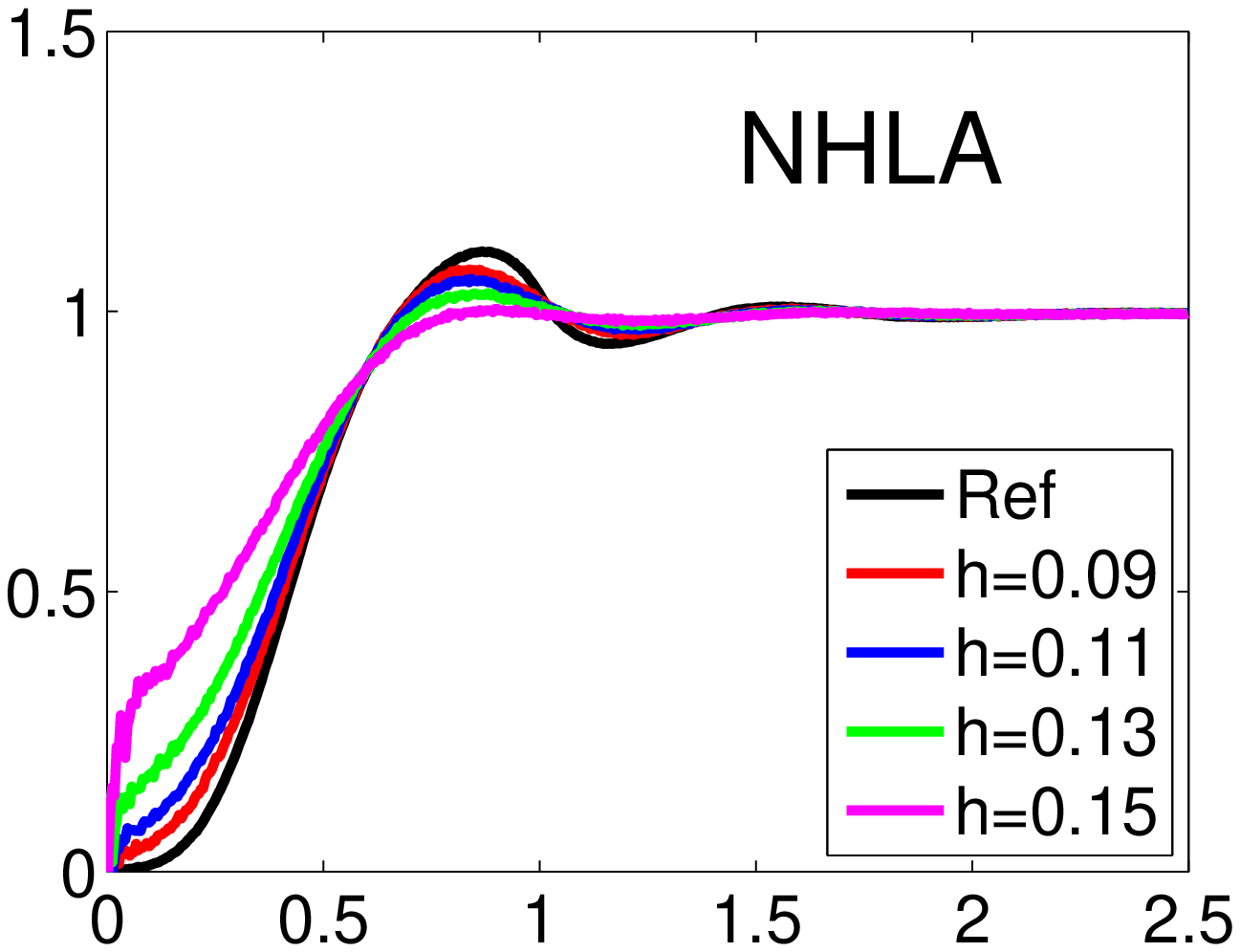}
\includegraphics[scale=0.3]{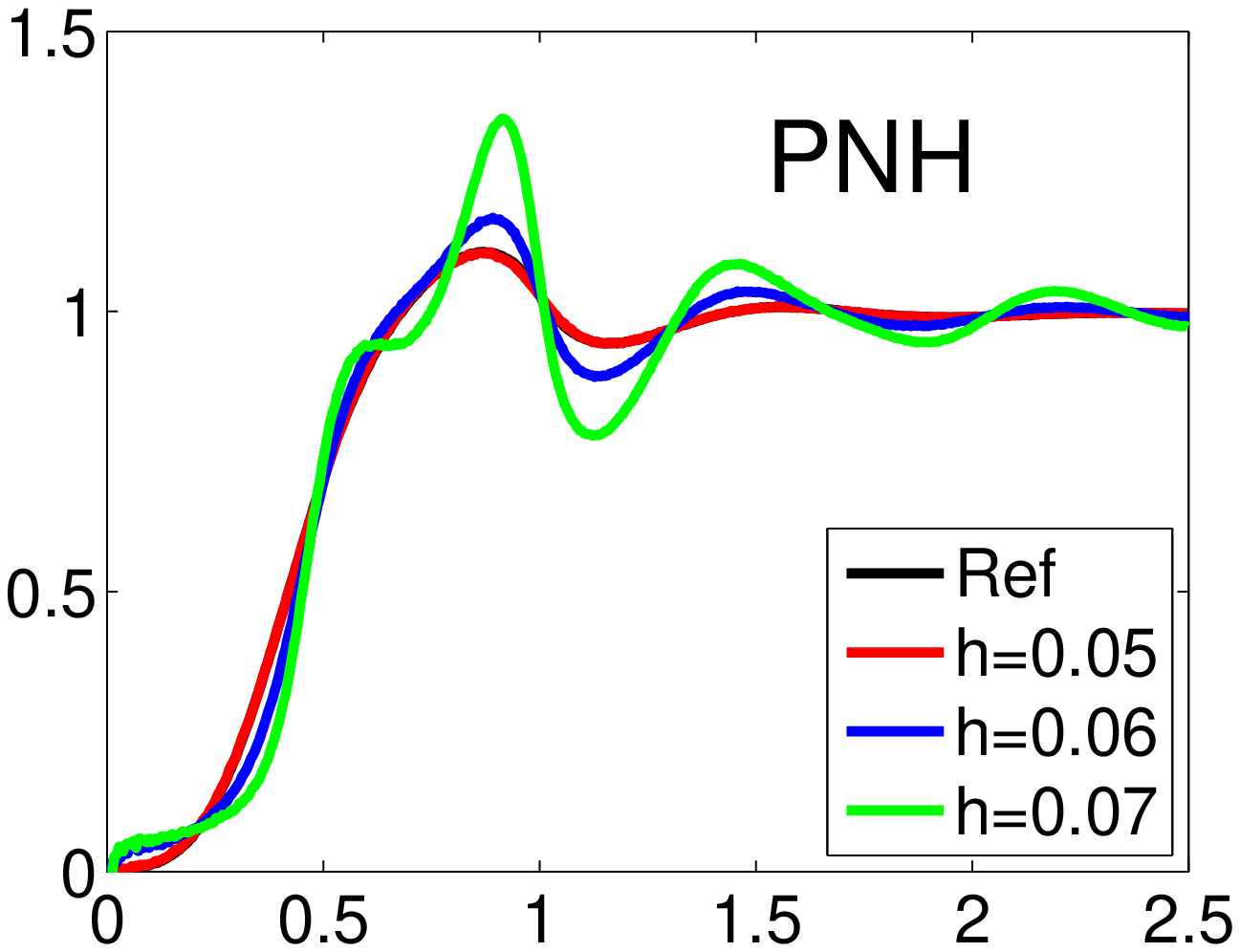}
\includegraphics[scale=0.3]{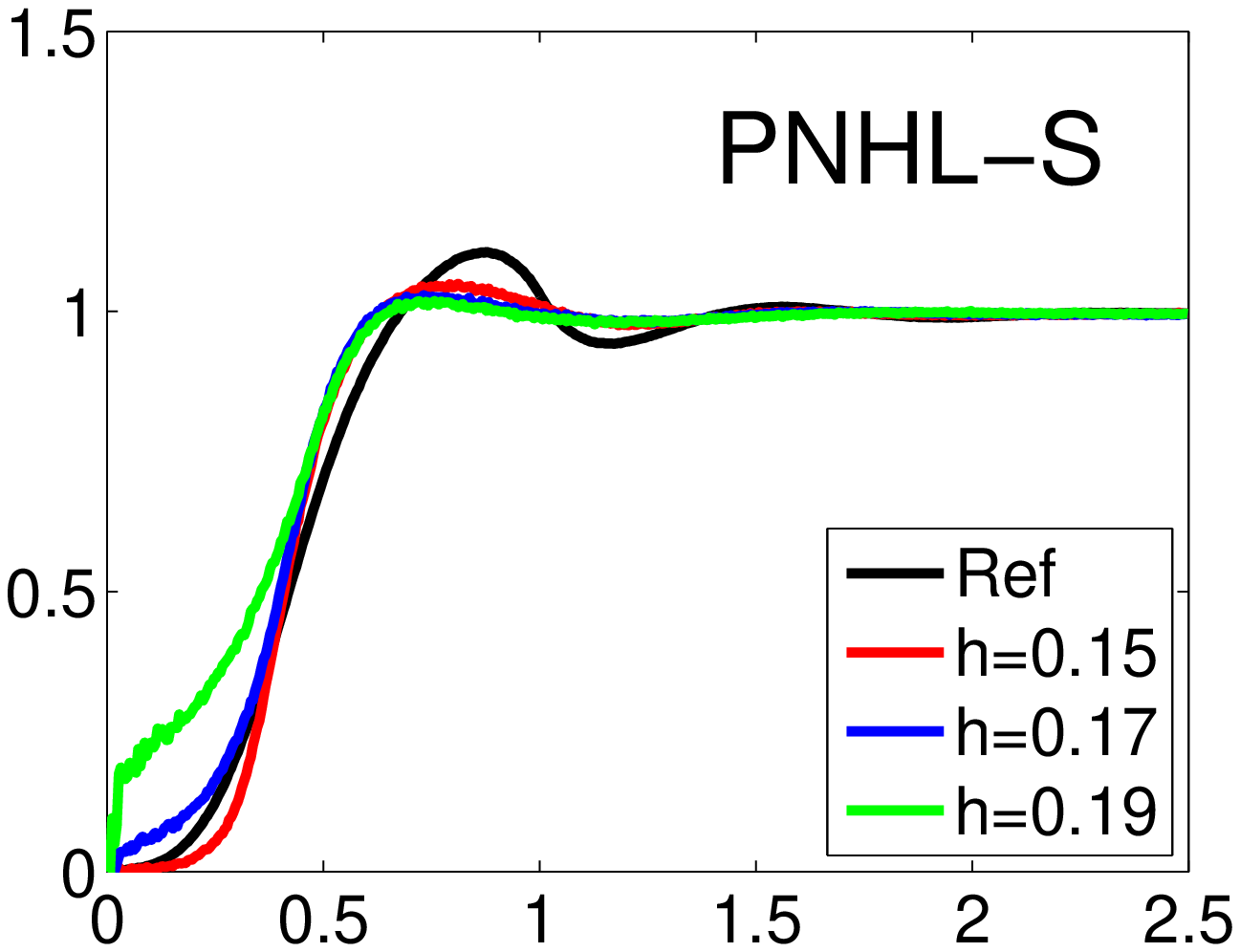}
\includegraphics[scale=0.3]{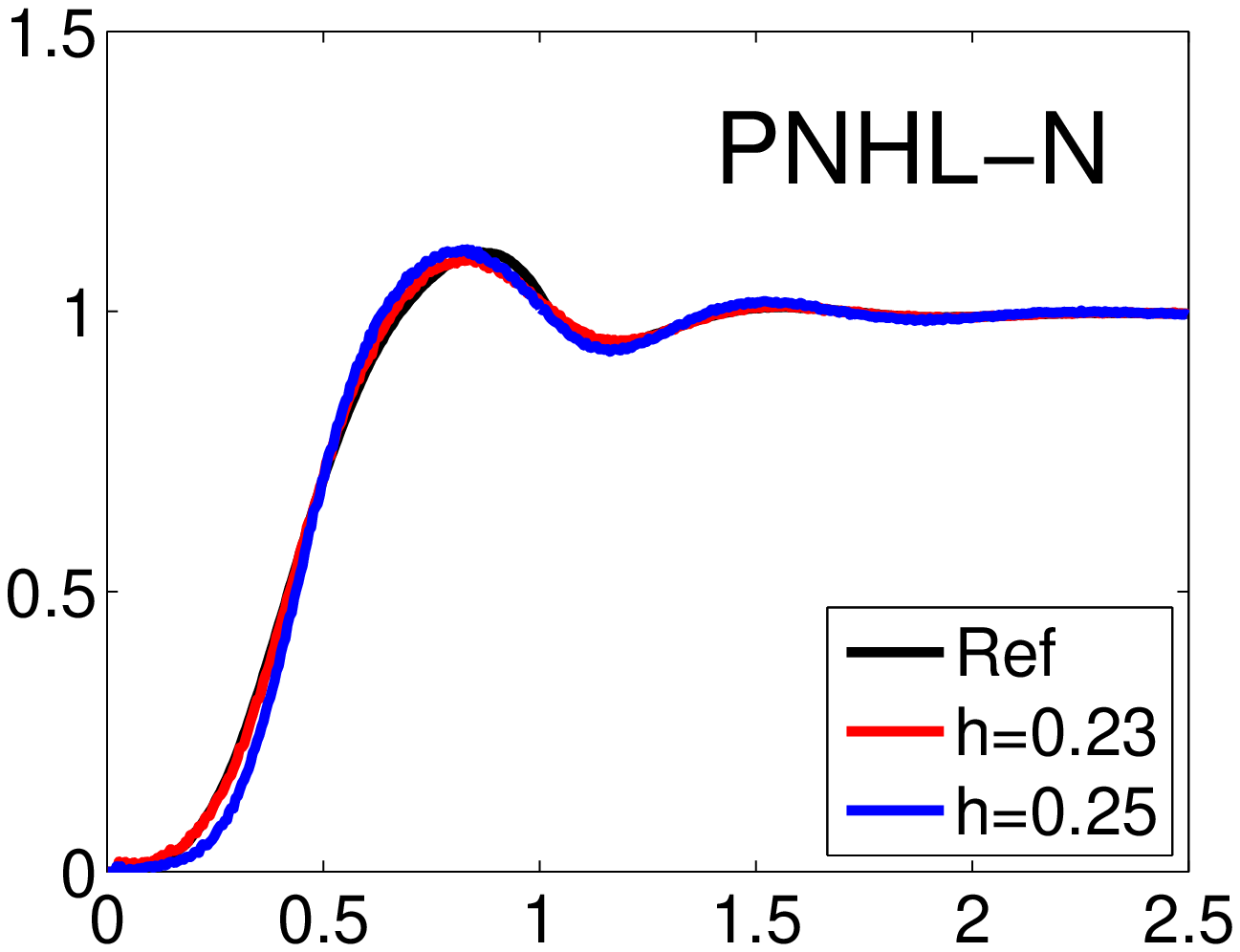}
\caption{\small (Color.) Stepsize effects on the radial distribution function (RDF) in various numerical methods were compared. The black solid lines are the reference solutions obtained using very small stepsizes $h=0.01$, while the colored lines correspond to different stepsizes. }
\label{fig:Comp_RDF_full}
\end{figure}

Fig. \ref{fig:Comp_RDF} compares the radial distribution function (RDF) that characterizes the structure of the fluids. The DPD-S1 and PNHL-N methods were used for the standard DPD and PNHL systems respectively. Given that very small stepsizes were used, all the methods display exactly the same RDF, which indicates that different systems maintain the same structure of the fluids without the impacts of discretization errors. Expectedly, discretization errors start to destroy the structure of the fluids with increasing stepsizes as shown in Fig. \ref{fig:Comp_RDF_full}. Standard DPD methods and Peters thermostat exhibit similar behaviors with the RDFs starting to show artifacts around $h=0.09$ and being heavily destroyed around $h=0.13$. The LA and PNH thermostats again show lower stability in maintaining the correct structure, blowing up around $h=0.11$ and $0.07$, respectively. The performance of PNHL-S and NHLA methods are slightly better than the majority of the schemes, while the PNHL-N method only starts to show pronounced artifacts above $h=0.25$, which is remarkably more than two times larger than the stepsize usable in the standard DPD methods.

\begin{figure}[htb]
\centering
\includegraphics[scale=0.5]{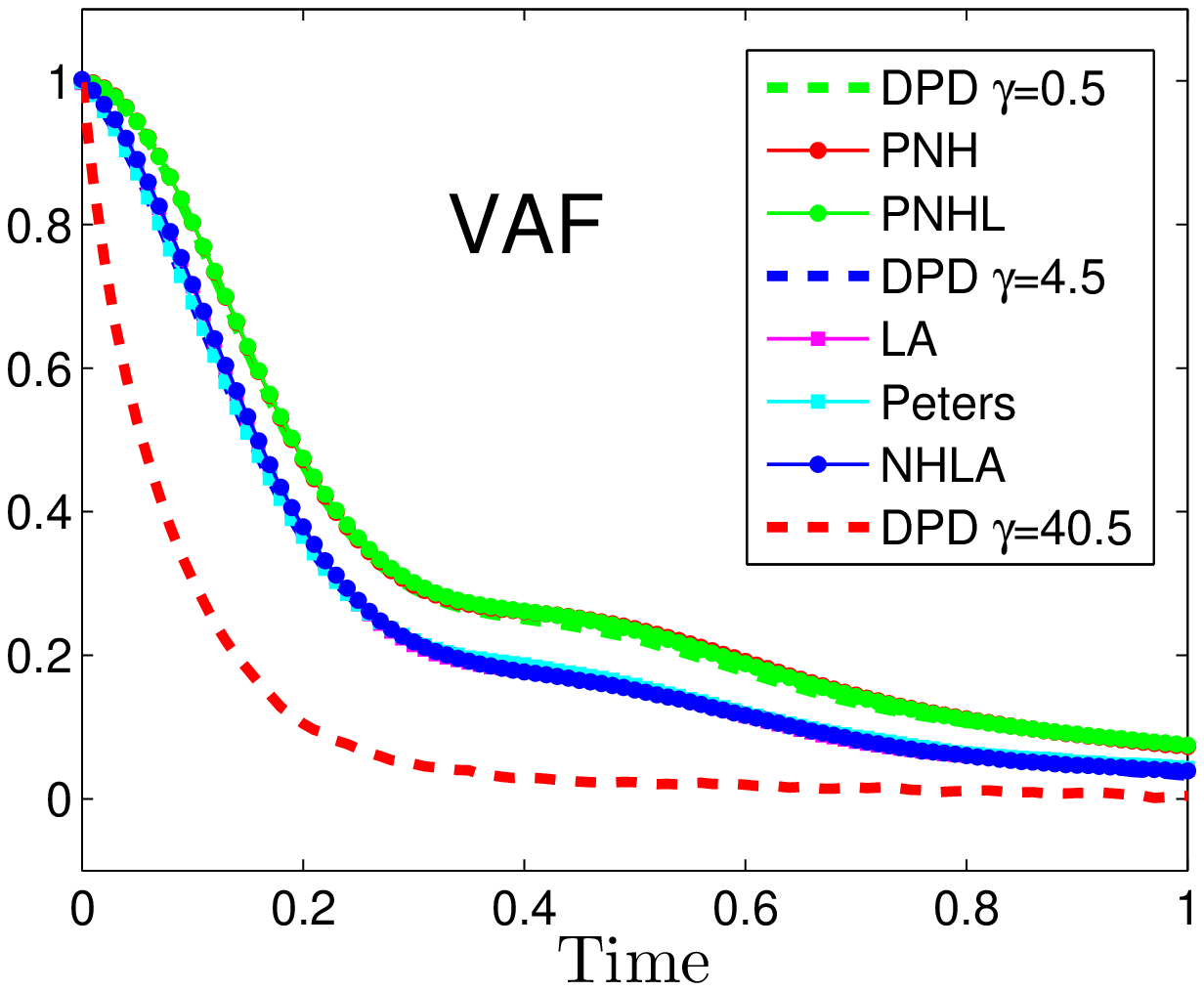}
\caption{\small (Color.) Comparisons of velocity autocorrelation function (VAF) of various numerical methods by using very small stepsize $h=0.01$. Standard DPD methods with three different values of friction coefficient were calculated by using DPD-S1 method to compare with other methods. 100 different runs were averaged to reduce the sampling errors after the system was well equilibrated.}
\label{fig:Comp_VAF}
\end{figure}

\begin{figure}[htb]
\centering
\includegraphics[scale=0.3]{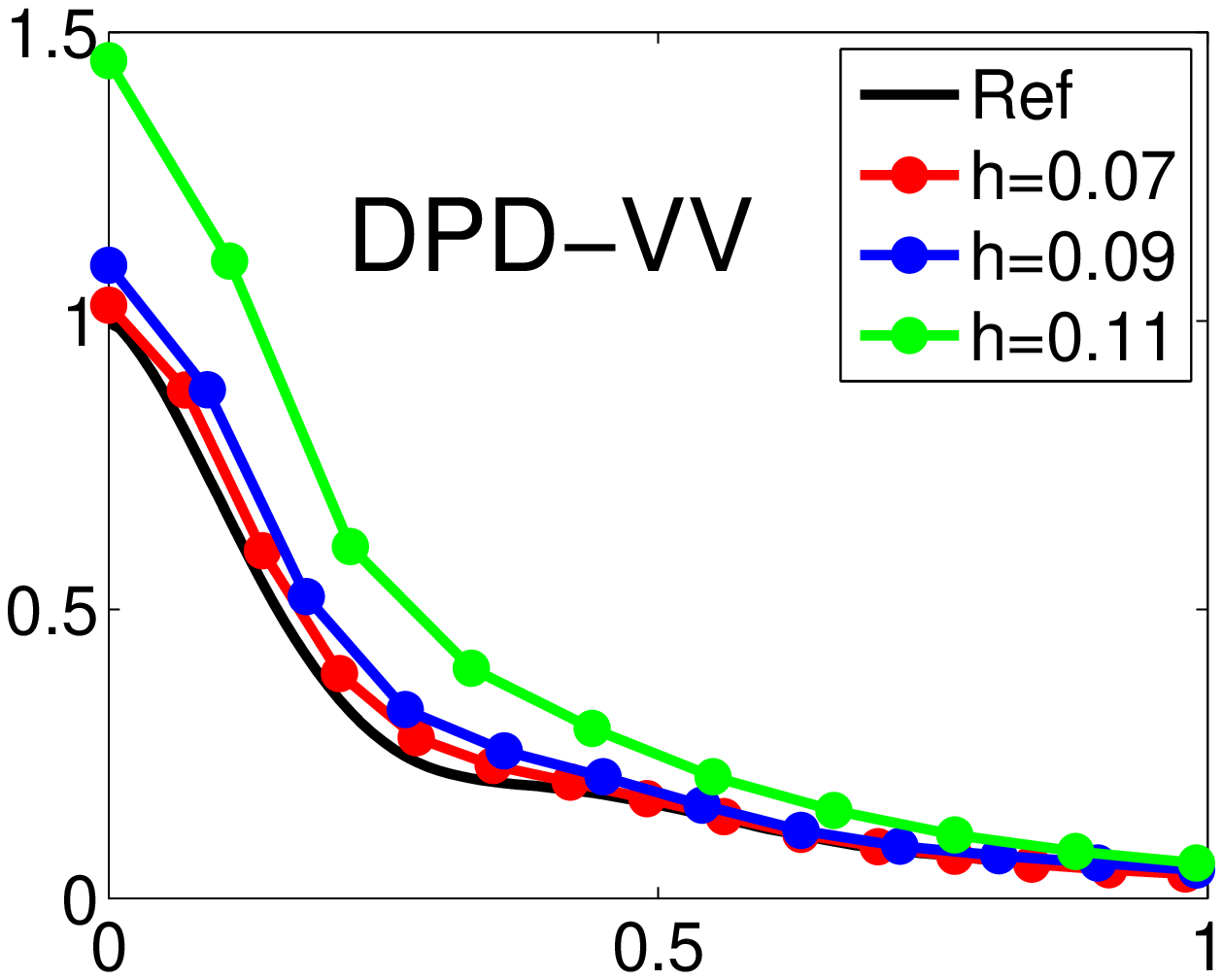}
\includegraphics[scale=0.3]{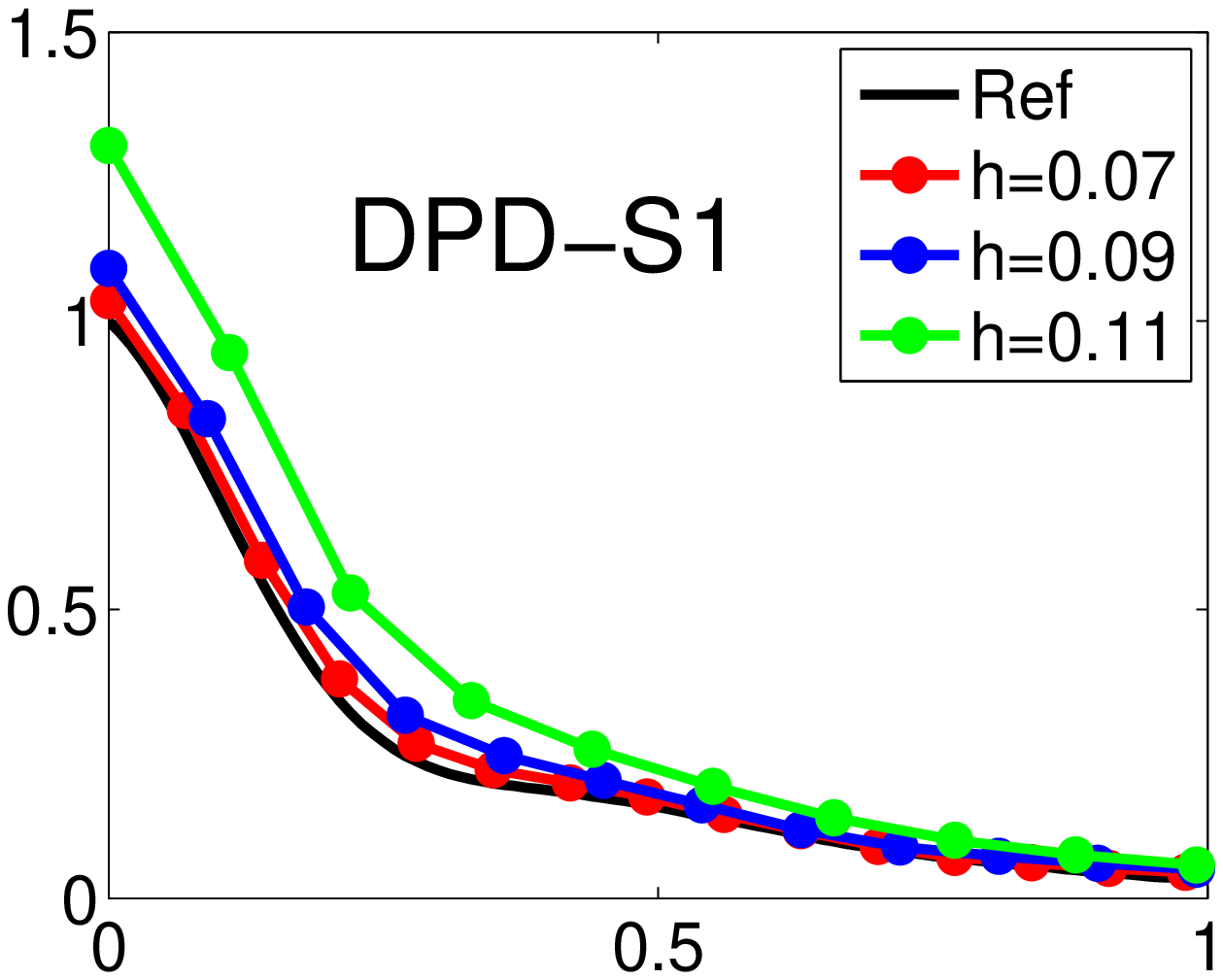}
\includegraphics[scale=0.3]{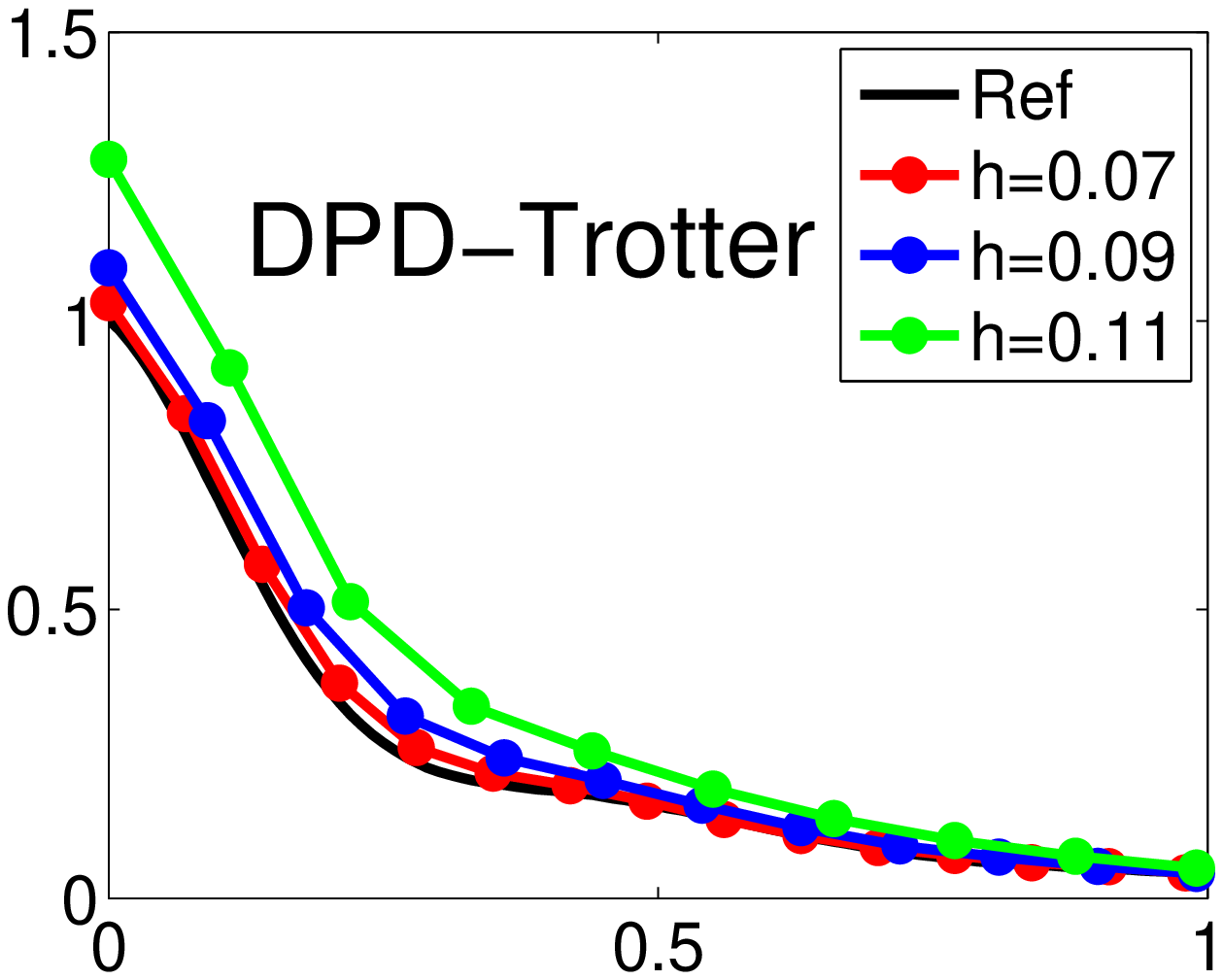}
\includegraphics[scale=0.3]{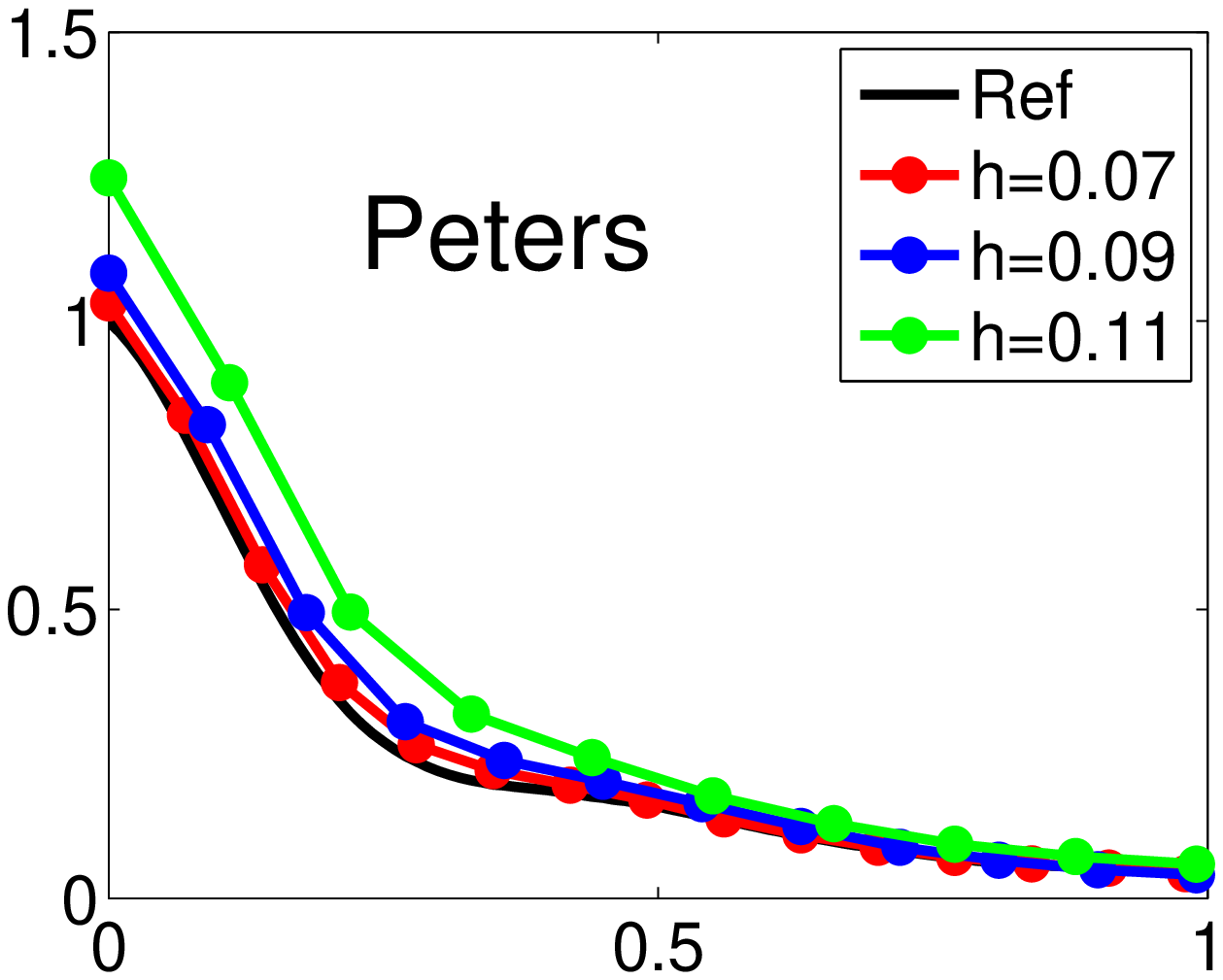}
\includegraphics[scale=0.3]{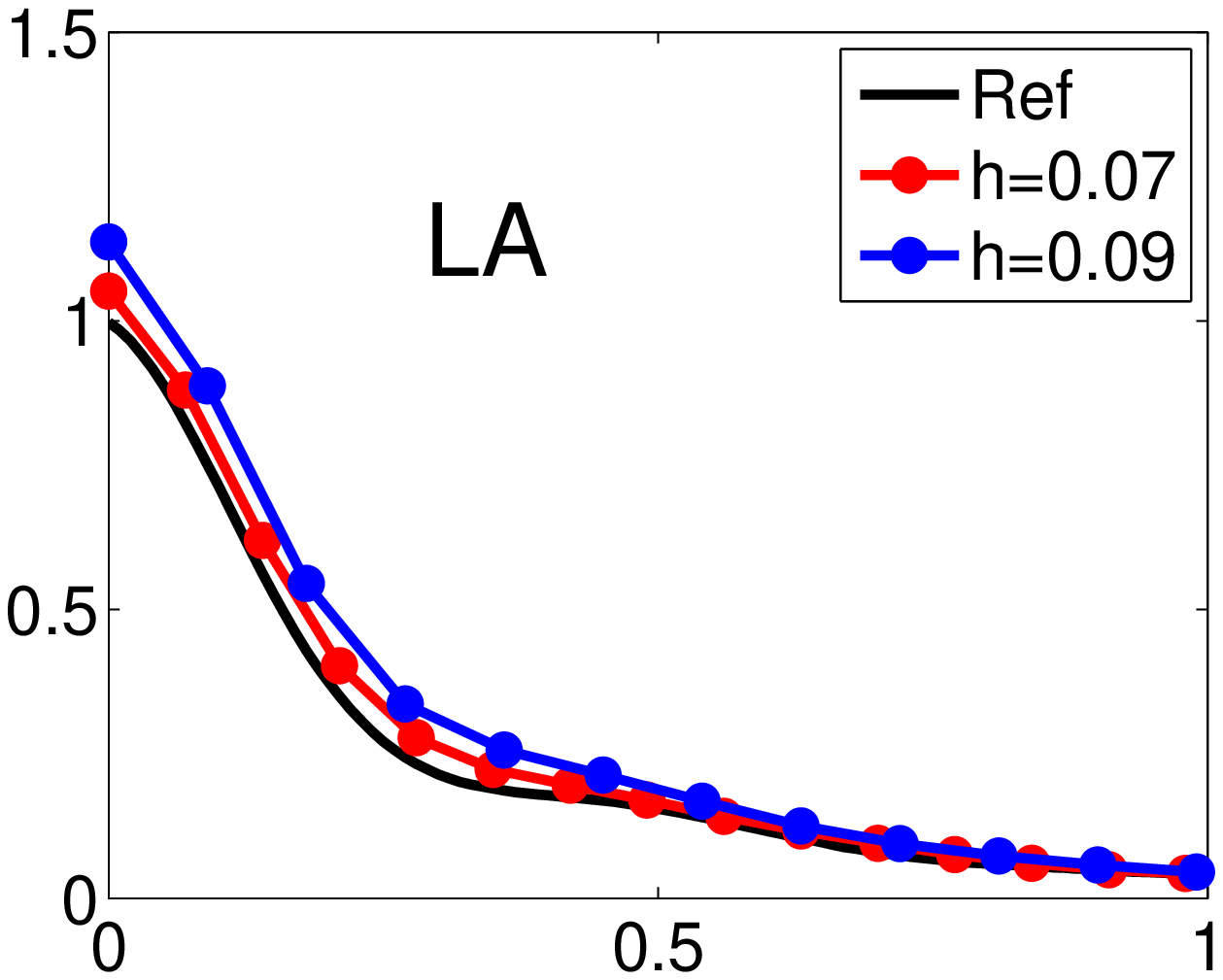}
\includegraphics[scale=0.3]{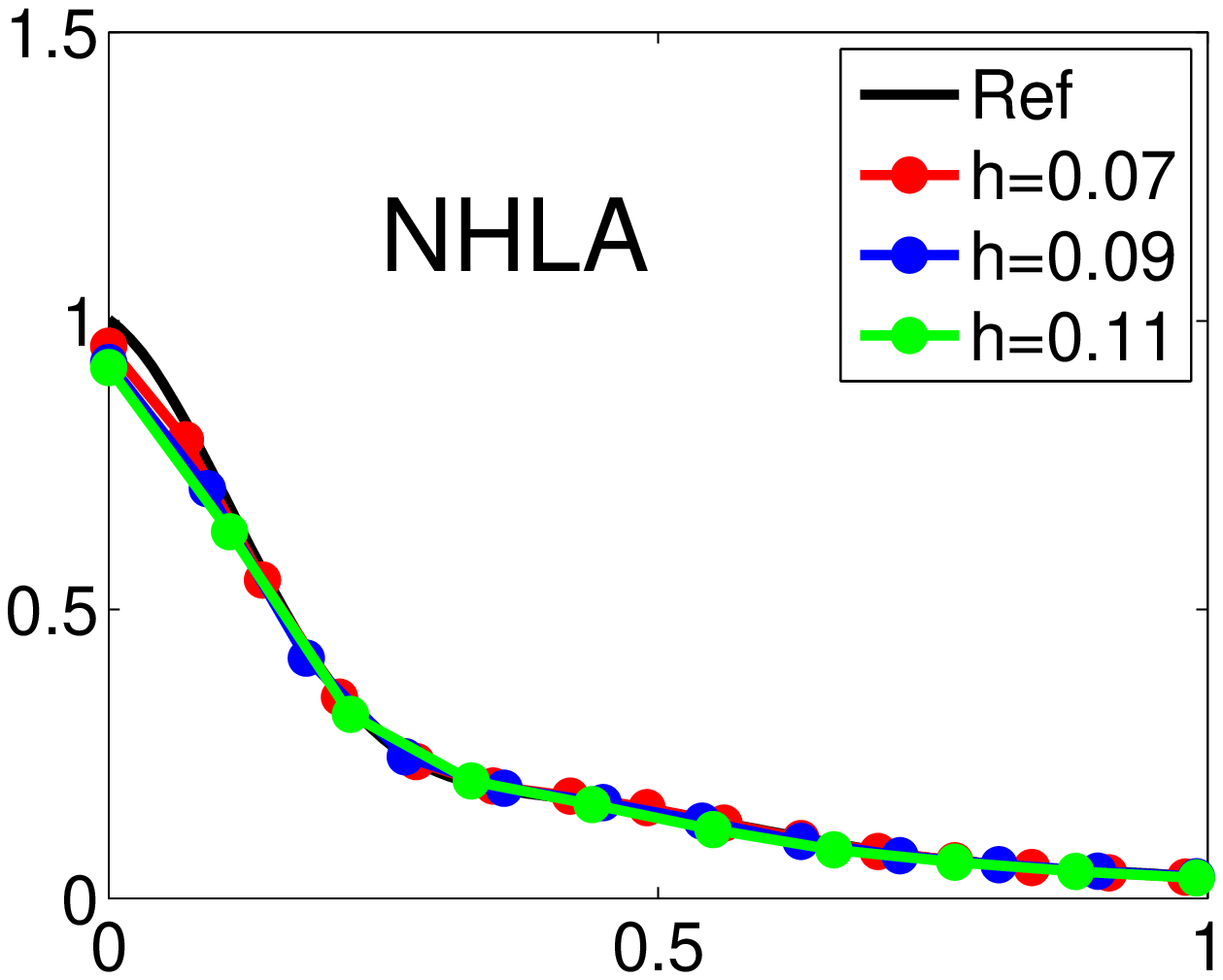}
\includegraphics[scale=0.3]{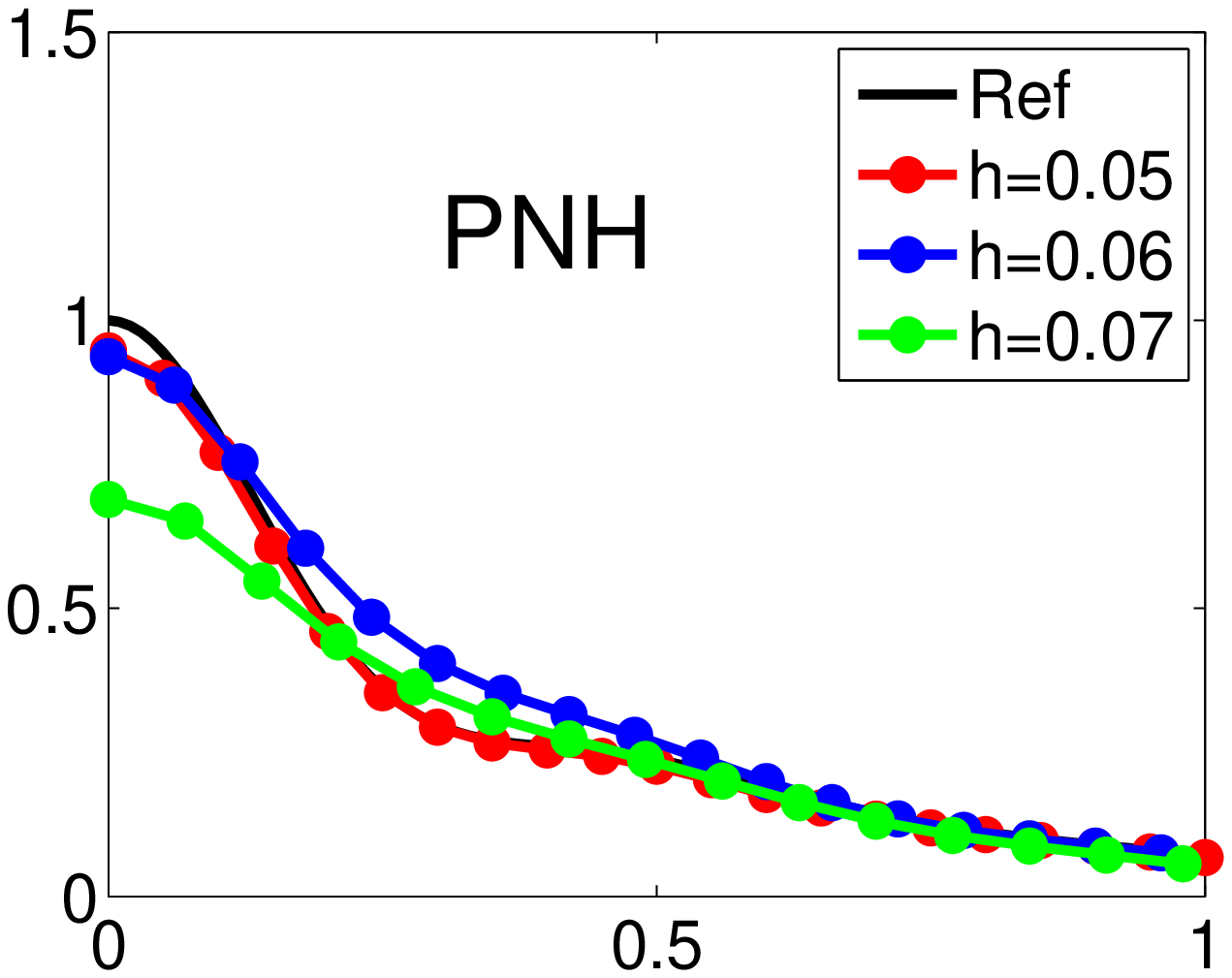}
\includegraphics[scale=0.3]{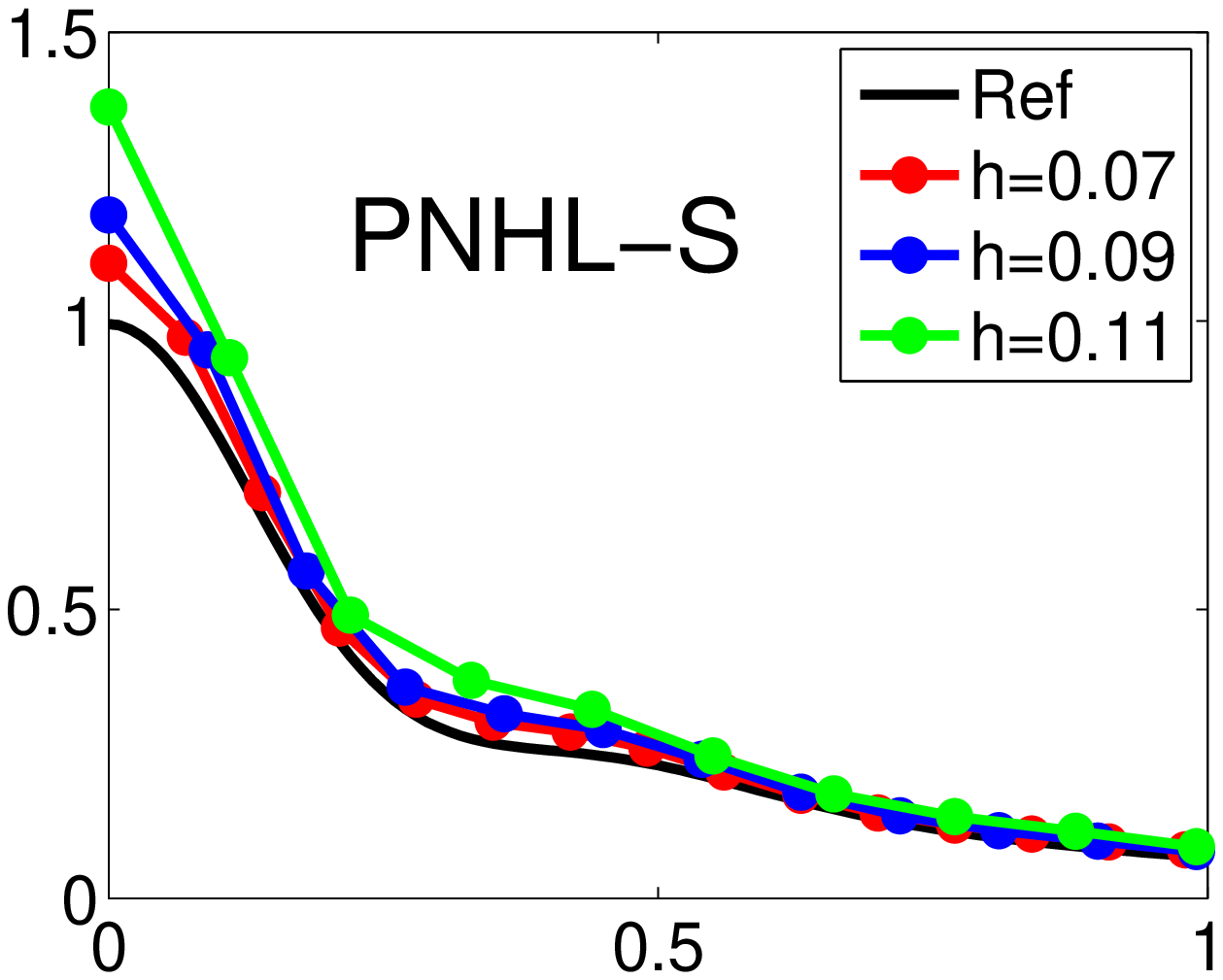}
\includegraphics[scale=0.3]{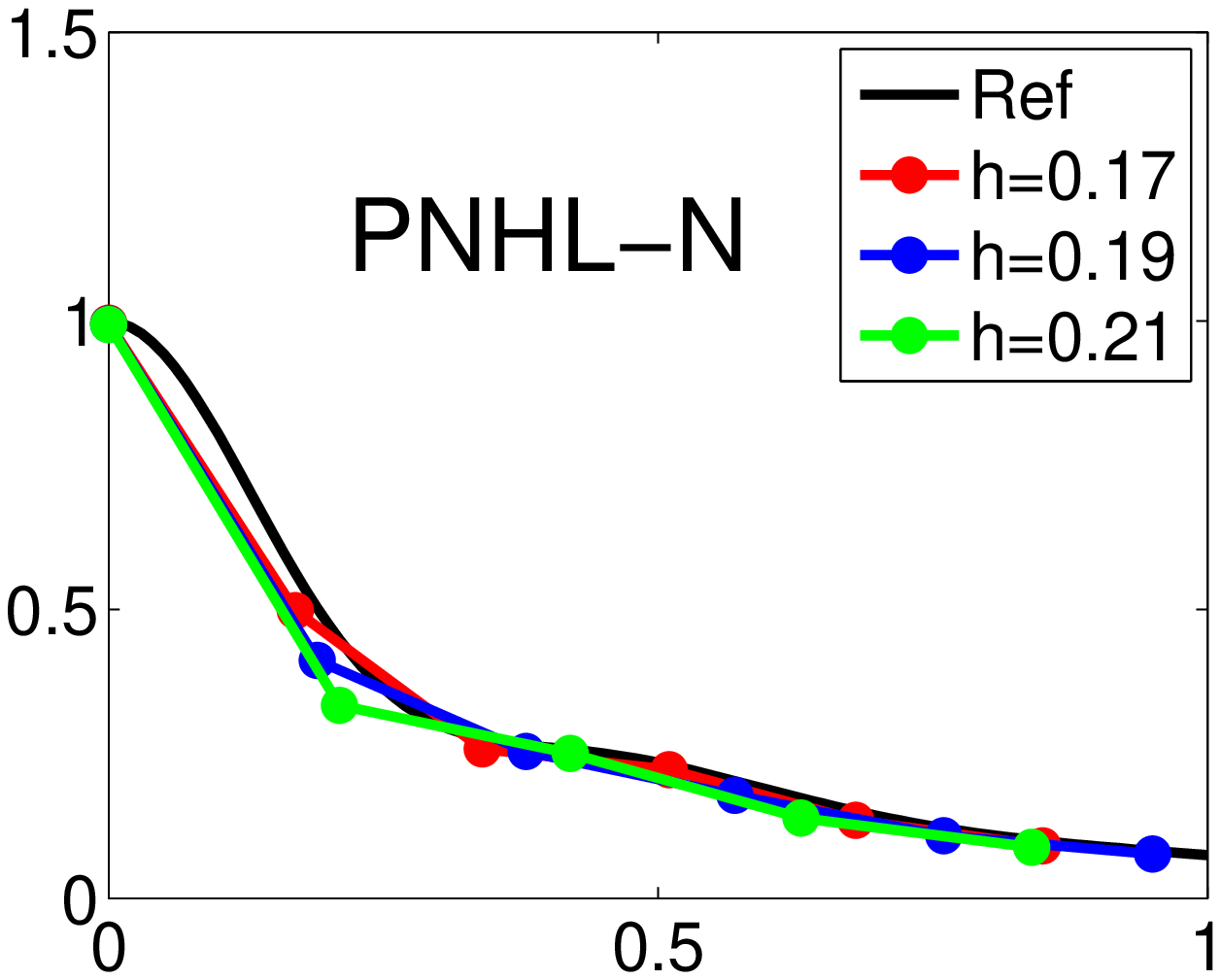}
\caption{\small (Color.) Stepsizes effects on the velocity autocorrelation function (VAF) in various numerical methods were compared. The black solid lines are the reference solutions obtained using very small stepsizes, while those colored lines correspond to different stepsizes. The error in the VAFs at time zero reflects the bias, visible at large stepsize, in the mean kinetic energy (which is the normalization factor). The VAFs could be rescaled so that they start from one, but it would mask the presence of this large disturbance from the target temperature.}
\label{fig:Comp_VAF_full}
\end{figure}

The velocity autocorrelation function (VAF) of various numerical methods were calculated in Fig. \ref{fig:Comp_VAF} to compare with standard DPD methods with three different values of friction coefficient. The DPD-S1 method was used to calculate the VAF of standard DPD methods since there is no difference with other two if very small stepsizes were used to reduce the effects of discretization errors. Similarly, the uncorrupted dynamics of PNHL-S and PNHL-N methods should be exactly the same, and the latter was used. The VAFs of the PNH and PNHL systems are indistinguishable and consistent with standard DPD methods in the regime of small friction coefficient $\gamma=0.5$. This is not surprising since the average of the dynamical variable $\xi$, which is controlled by an additional thermostat and can be thought of as the ``dynamical friction'', tends to zero.  As we expected, the Peters thermostat shares the same VAF as standard DPD methods in the regime of the commonly-used friction coefficient $\gamma=4.5$ if the same value of $\gamma$ was chosen. The stochastic randomization frequency $\Gamma=0.44$ was used in LA and NHLA thermostats to maintain similar translational diffusion properties as in standard DPD system with $\gamma=4.5$. The VAF of standard DPD system with large friction coefficient $\gamma=40.5$ is also shown in the figure, which indicates that the larger the friction is the faster the VAF goes down (the system loses memory faster).

\begin{figure}[tb]
\centering
\includegraphics[scale=0.5]{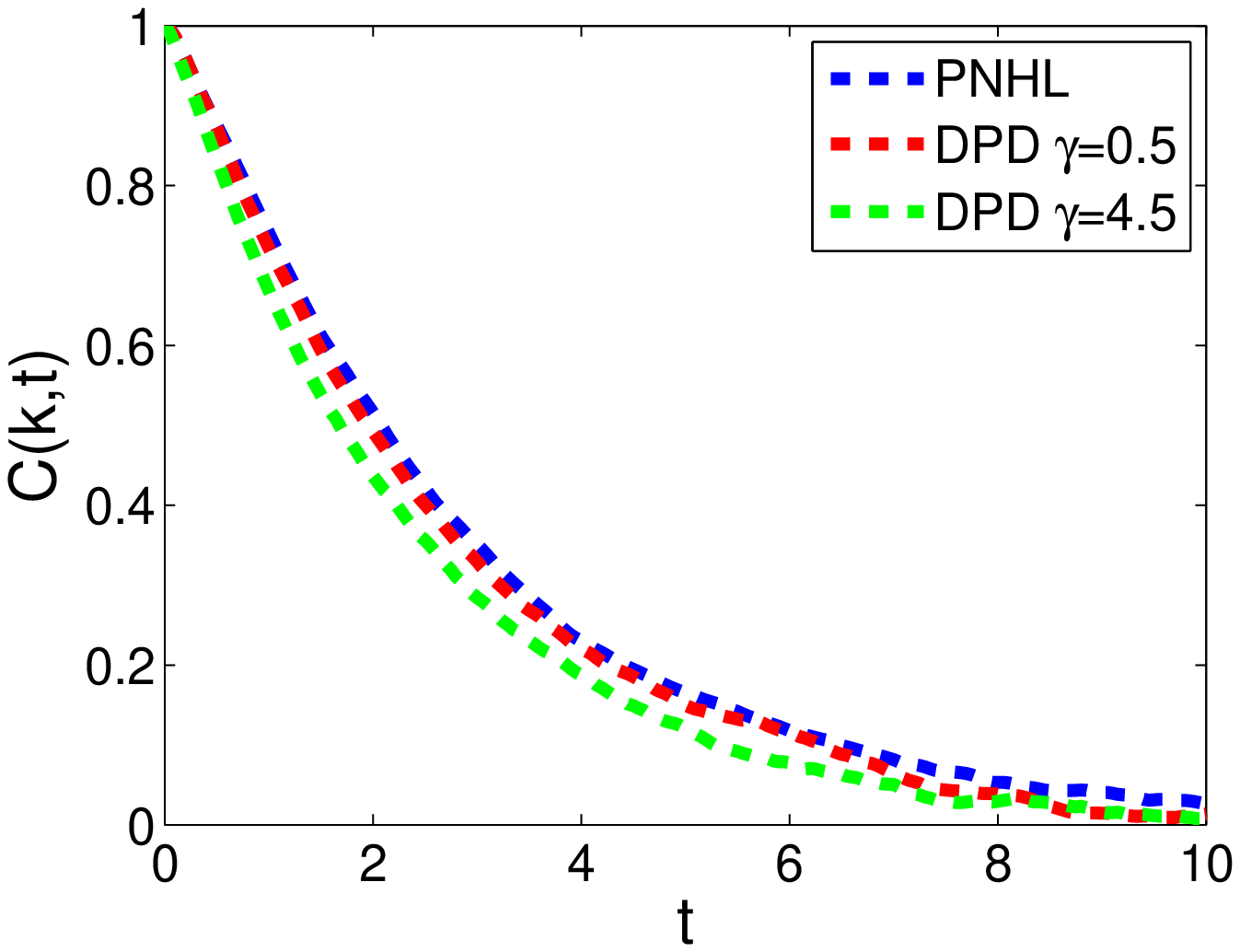}
\includegraphics[scale=0.5]{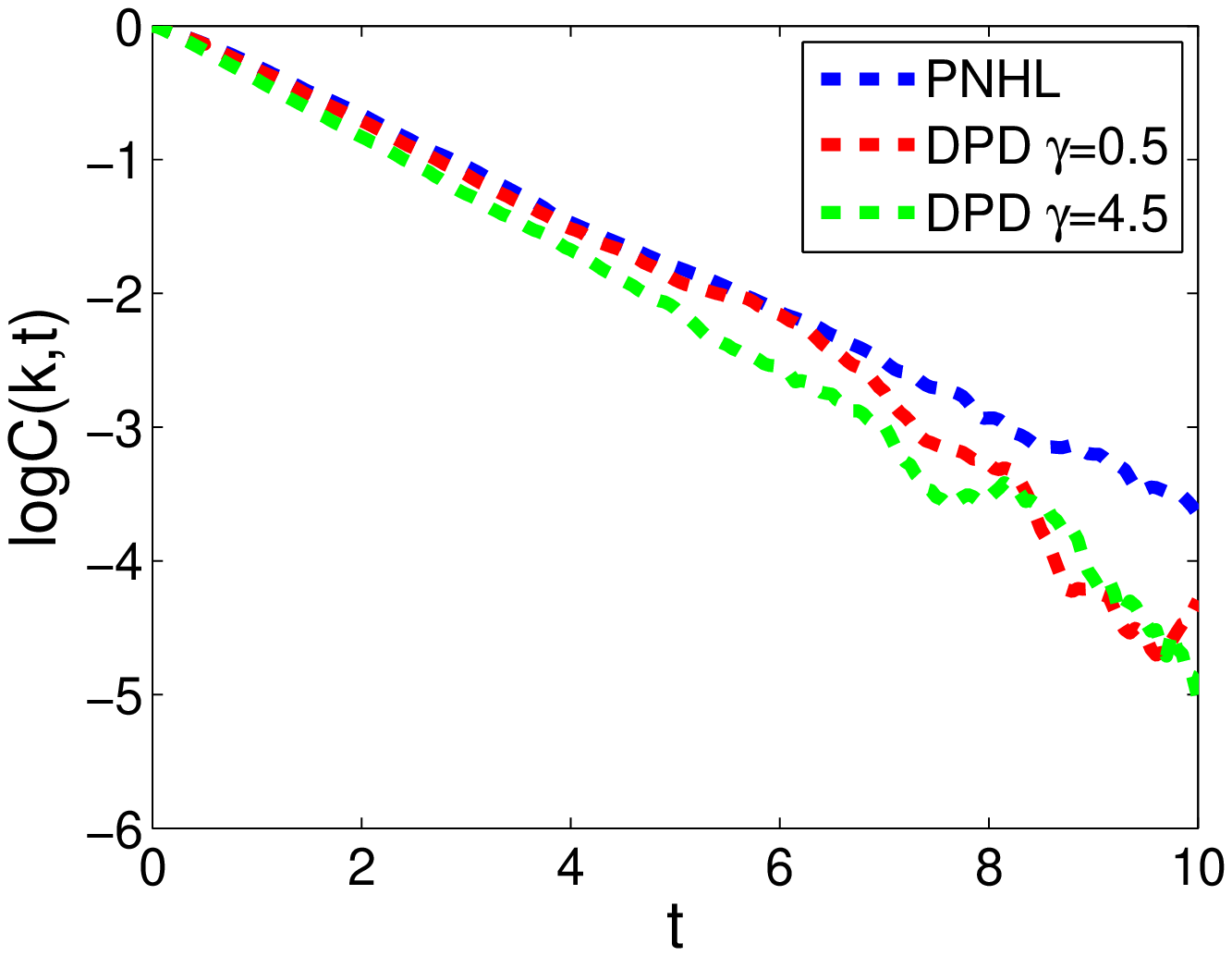}
\caption{\small (Color) Comparisons of the normalized transverse momentum autocorrelation function (left) and its logarithm (right) between standard DPD and PNHL methods with stepsize $h=0.05$. DPD-S1 and PNHL-N were used to solve the DPD and PNHL systems respectively. The ratio of the curves on the right panel is proportional to the corresponding shear viscosity. The wavenumber $k$ was chosen as $2\pi/L$ and 100,000 different runs were averaged to reduce the sampling errors after the system was well equilibrated.}
\label{fig:Comp_TMAF_and Log}
\end{figure}

The effects of the discretization error on each method were also investigated in Fig. \ref{fig:Comp_VAF_full}. The results are largely consistent with those observed for the configurational temperature. The only surprise is that the PNHL-S method allows a maximum stepsize which is similar to those of the thermostats considered (and much lower than the useful stepsize for PNHL-N). Moreover, we observe that the PNH thermostat begins to display non-physical artifacts, at a stepsize of just $h=0.05$. Likewise, the LA thermostat has a lower stability threshold. Among the various schemes, PNHL-N is again by far the most stable scheme, exhibiting only a mild deviation from the reference VAF at $h=0.17$.

One of the most important features of DPD simulations is the correct handling of rotational relaxation which will be important for resolving correct vortical motion and long-ranged interactions.  We therefore investigate the computation of transverse momentum autocorrelation function (TMAF, see Section \ref{Subsubsection:Dynamical}) for each scheme.  The results here are in many ways similar to those obtained for the velocity autocorrelation function (Fig. \ref{fig:Comp_VAF}), thus only the results of DPD and PNHL methods are shown in Fig. \ref{fig:Comp_TMAF_and Log}: the PNHL method is expectedly consistent with small friction ($\gamma =0.5$) in standard DPD simulation, and only a minor difference is observed between the PNHL method and standard DPD with a moderate value of friction ($\gamma = 4.5$). One may notice that the TMAF shown in Fig. \ref{fig:Comp_TMAF_and Log} is not as smooth as the VAF in Fig. \ref{fig:Comp_VAF} even by averaging 1000 times more different runs. We emphasize here that collective ($N$-particle) correlations, such as TMAF and stress autocorrelation function \cite{Chaudhri2010}, fluctuate rapidly and thus are always determined with poorer statistics than single-particle ones, such as VAF.

We emphasize the following facts regarding our numerical tests:
\begin{itemize}
  \item
    A detailed statistical analysis of the results presented has not been incorporated in this article, due to the extensive computational requirements of doing so. An early study \cite{Shardlow2003} has already suggested that highly reliable estimation of the kinetic temperature can be typically obtained by various methods in the stepsize regime of our interests, i.e. $h \geq 0.05$. In terms of convergence rate of thermodynamic properties to distribution, all the methods perform similarly in practice (see Figs. \ref{fig:Comp_KT}-\ref{fig:Comp_CT_U}).
  \item
    The influence of the friction coefficients (three different values, $\gamma=$ 0.5, 4.5 and 40.5), both on the maximal timestep and the dynamics of the system, has been investigated in this section. Different values of the friction coefficients gave little difference in the control of those static quantities we calculated, thus only the results of $\gamma=4.5$ have been presented. However, the dynamical properties of the system did appear to depend on the strength of the friction as shown in the velocity and transverse momentum autocorrelation functions.
  \item
    Based on the results for the VAF and TMAF, we have observed very good agreement of the dynamical properties between PNHL and DPD with relatively small friction coefficient, particularly $\gamma=0.5$. Although the PNHL method may not be able to fully recover the hydrodynamics as DPD, we have already seen that PNHL offers substantially improved stability in simulations. Moreover, the PNHL method, as a general momentum-conserving thermostat, has a valid motivation in a broader context than just comparison with DPD; in particular it will be useful in NEMD and in other cases where fluid dynamics per se is not at issue.
\end{itemize}

\subsection{Computational Efficiency}

The computational efficiency of the various methods was tested with simulation details in Section \ref{Subsection:Simulation_Details}. All the tests were run on an HP Z600 Workstation with 15.7 GB RAM. As shown in Table \ref{table:efficiency}, we calculated the CPU time (milliseconds) taken with the use of Verlet neighbor lists for the integration of a single time step of $h=0.05$ (averaged over 10000 consecutive time steps). Note that only the time for the main integration step (without calculating any physical quantity) was counted.

\begin{table}[htbp]
\begin{center}
\resizebox{1.0\textwidth}{!}{\begin{minipage}{\textwidth}
    \begin{tabular}{ |c|C{2.3cm}|C{2.3cm}|C{2.6cm}|C{1.7cm}|C{2.3cm}| }
    \hline
    \textbf{Method} & \textbf{Critical Stepsize} & \textbf{Maximal Stepsize} & \textbf{Force Calculation} & \textbf{CPU Time} & \textbf{Scaled Efficiency} \\ \hline
    DPD-VV      & 0.05 & 0.10 & 1 & 17.710 & 100.0\% \\ \hline
    DPD-S1      & 0.05 & 0.11 & 1 & 17.783 & 99.6\% \\ \hline
    DPD-Trotter & 0.05 & 0.11 & 1 & 18.482  & 95.8\% \\ \hline
    LA          & 0.05 & 0.10 & 1 & 15.364 & 115.3\% \\ \hline
    Peters      & 0.05 & 0.11 & 1 & 18.286 & 96.9\% \\ \hline
    NHLA        & 0.07 & 0.13 & 1 & 16.281 & 152.3\% \\ \hline
    PNH         & 0.05 & 0.08 & 1 & 13.990 & 126.6\% \\ \hline
    PNHL-S      & 0.08 & 0.17 & 1 & 19.408 & 146.0\% \\ \hline
    PNHL-N      & 0.17 & 0.23 & 2 & 32.183 & 187.1\% \\
    \hline
    \end{tabular}
\caption[Table caption text]{Comparisons of the computational efficiency of the various numerical methods. ``Critical stepsize'' is the stepsize beyond which the numerical method starts to show pronounced artifacts, while ``maximal stepsize'' is the stepsize stability threshold. The ``numerical efficiency'' of each method was scaled to that of the benchmark DPD-VV method.}
\label{table:efficiency}
\end{minipage} }
\end{center}
\end{table}

In order to quantitatively compare the overall performance of each method, we define a quantity, the``numerical efficiency'', that measures the amount of simulation time accessible per unit of computational work, i.e.
\begin{equation*}
  \text{Numerical Efficiency} = \frac{\text{Critical Stepsize}}{\text{CPU Time Per Step}},
\end{equation*}
where the ``critical stepsize'' is defined as the stepsize beyond which pronounced artifacts become apparent in some observable thus rendering the simulation unusable.    For practical purposes, we use a threshold of 10\% relative error in configurational temperature in determining the critical stepsize.  Our reasoning in choosing the configurational temperature as the quantity for determination of the critical stepsize is that (a) it is a sensitive observable and difficult to control in simulation, (b) accuracy of configurational temperature seems to us to be intuitively to be a core requirement for canonical simulation methods, (c) good control of the configurational temperature appears to imply good results in other stationary computations.   The choice of 10\% as the bar for accuracy is clearly arbitrary.   If a smaller error threshold were used, the results may change slightly, but the ordering of the methods in terms of efficiency would remain essentially the same.

We also show the ``maximal stepsize'' in Table \ref{table:efficiency}, which is the stepsize at which the numerical method is either above 100\% relative error in configurational temperature or unstable. Only the PNHL-N method needs to calculate the force twice in each integration step, which is the reason why an almost doubled time was needed than other methods. Finally, the computed ``numerical efficiency'' of each method was scaled to that of the DPD-VV method since it is still the most popular method in DPD simulations and we use that method as the benchmark in the comparisons.

As can be seen from the table, those standard DPD methods and the Peters thermostat have comparable ``numerical efficiency''. The LA and PNH thermostats are slightly better than the commonly-used DPD-VV method with around 15\% and 27\% improvements, respectively. Both NHLA and PNHL-S methods maintain about 50\% enhancement. The improvement of the ``numerical efficiency'' of the PNHL-N is remarkably 87\% in comparison to the DPD-VV method. Similarly, we may also compute the enhancements of the PNHL-N method based on kinetic temperature and average potential energy, using the same critical stepsize $h=0.05$ as in configurational temperature: 43\% for the former and 120\% for the latter.

\section{Conclusions}
\label{Section:Conclusions}

We have reviewed a number of numerical methods that are widely used in DPD simulations and have also proposed a new stochastic momentum-conserving thermostat, pairwise Nos\'{e}-Hoover-Langevin (PNHL) thermostat. Two favorable splitting methods of the PNHL thermostat were introduced and compared with existing methods in the computation of various physical quantities (both static and dynamical).

We have observed that, for the PNHL thermostat proposed here, the PNHL-N method based on a nonsymmetric arrangement of the terms of a splitting, gives an enormous stability benefit.  The PNHL-N method needs to calculate the force twice in each integration step, which is computationally costly in the model setting used in this article; nevertheless when the computational overhead is costed carefully the PNHL-N method outperforms the alternatives. To measure the practical performance of numerical methods in DPD simulations quantitatively, we have defined the ``numerical efficiency'', based on which we have reported substantially improvements of both methods of the newly proposed thermostat, with the symmetric PNHL-S method 46\% more efficient than the commonly-used DPD-VV method and the nonsymmetric PNHL-N method incredibly 87\% better than the benchmark method in DPD simulations. It should be noted that, if the force calculation is not that expensive in other model settings, the gain of using the PNHL-N method could be further exploited.

Based on the numerical experiments of the velocity and transverse momentum autocorrelation functions which characterize the translational and rotational diffusions of the system respectively, DPD and PNHL give rather similar dynamical properties in practice.  Although PNHL as formulated is not based on a hydrodynamic interaction model \cite{Kinjo2007,Hijon2010}, we have seen that it is an effective replacement for DPD in the low-friction regime. Moreover, we point out that the projection along interacting particle pairs in PNHL could be replaced by alternatives to achieve further control of transport properties. The method is also potentially useful more broadly in molecular simulation applications, whenever momentum conservation is at issue.

The only difference between PNH and PNHL is the additional Langevin thermostat acting on the dynamical friction, however they perform very differently.  This is probably because the cutoffs used in PNH and the simple potentials there provided insufficient internal mechanisms for the system to achieve an ergodic sampling of the canonical distribution. In molecular dynamics with steep pair potentials (e.g. Lennard-Jones), the ergodic properties develop more naturally and the ``L'' in PNHL can be redundant in some instances.  It is also worth mentioning that both PNH and PNHL are not able to vary the Schmidt number since the average of the dynamical friction tends to zero, whereas the Schmidt number can be tuned in some of the other schemes.

We have also investigated the order of convergence of the long-time averages to the invariant measure for a couple of methods described in this article. By extending the framework recently introduced in Langevin dynamics, we can infer (and verify using numerics) the second order convergence for those nonsymmetric GLA-like methods (DPD-S1, LA and Peters thermostats). However, rigorous investigation on other nonsymmetric methods (DPD-VV, NHLA, PNH and PNHL-N methods) that surprisingly obtained second order convergence remains to be established. Overall, we claim here that PNHL thermostat indeed can be used (and may be preferred in some typical cases) as an alternative to low-friction DPD simulations with substantially improved computational efficiency and no degradation of convergence rate.

\section*{Acknowledgements}

The authors thank Charles Matthews for stimulating discussions. The authors further thank Michael Allen, Gabriel Stoltz and anonymous referees for valuable suggestions and comments. BL acknowledges the support of the Engineering and Physical Sciences Research Council (UK) and grant EP/G036136/1. XS gratefully acknowledges the financial support from the
University of Edinburgh and China Scholarship Council.


\appendix




\begin{appendices}
  \renewcommand\thetable{\thesection\arabic{table}}
  \renewcommand\thefigure{\thesection\arabic{figure}}

  \section{Order of Long-time Error} \label{Section:Appendix_Order}

To investigate the order of convergence of the invariant distribution, we define the operator $\hat{\mathcal{L}}^{\dag}$ associated with propagation under the numerical method with stepsize $h$ for an SDE.   It can be thought of as a perturbation of the exact Fokker-Planck operator $\mathcal{L}^{\dag}_\text{Exact}$
\begin{equation}\label{eqn:operator_perturbation}
  \hat{\mathcal{L}}^{\dag} = \mathcal{L}^{\dag}_\text{Exact} + h\mathcal{L}^{\dag}_{1} + h^{2}\mathcal{L}^{\dag}_{2} + O(h^{3}),
\end{equation}
for some perturbation operators $\mathcal{L}^{\dag}_{i}$.

We also view the invariant distribution $\hat{\rho}$ associated with the numerical method as a perturbation of the target canonical distribution $\rho_{\beta}$
\begin{equation}\label{eqn:distribution_perturbation}
  \hat{\rho} = \rho_{\beta}[1+hf_{1}+h^{2}f_{2}+h^{3}f_{3}+O(h^{4})],
\end{equation}
for some correction function $f_{i}$ satisfying $\langle f_{i} \rangle=0$.

Substituting $\hat{\mathcal{L}}^{\dag}$ and $\hat{\rho}$ into the stationary Fokker-Planck equation
\begin{equation*}
  \hat{\mathcal{L}}^{\dag}\hat{\rho} = 0
\end{equation*}
gives
\begin{equation*}
  \left(\mathcal{L}^{\dag}_\text{Exact} + h\mathcal{L}^{\dag}_{1} + h^{2}\mathcal{L}^{\dag}_{2} + O(h^{3})\right)\left(\rho_{\beta}[1+hf_{1}+h^{2}f_{2}+h^{3}f_{3}+O(h^{4})]\right)=0.
\end{equation*}
Since the exact operator preserves the canonical distribution, i.e. $\mathcal{L}^{\dag}_\text{Exact}\rho_{\beta}=0$, we obtain
\begin{equation}\label{eqn:error_analysis_PDE}
  \mathcal{L}^{\dag}_\text{Exact}(\rho_{\beta}f_{1}) = -\mathcal{L}^{\dag}_{1}\rho_{\beta},
\end{equation}
by equating first order terms in $h$.

For all the splitting schemes we are able to find the perturbation operator $\mathcal{L}^{\dag}_{1}$ by using the Baker-Campbell-Hausdorff (BCH) expansion.  Formally we can calculate its action on $\rho_{\beta}$, then the leading order correction function $f_{1}$ would be the solution of the partial differential equation (\ref{eqn:error_analysis_PDE}).

According to the definition of the BCH expansion, for linear operators X and Y, we have
\begin{equation*}
  e^{hX}e^{hY} = e^{h( X+Y + \frac{h}{2}[X,Y] + \frac{h^{2}}{12}([X,[X,Y]]-[Y,[X,Y]]) + O(h^{3}) )},
\end{equation*}
and
\begin{equation*}
  e^{\frac{h}{2}X}e^{hY}e^{\frac{h}{2}X} = e^{h( X+Y + \frac{h^{2}}{12}([Y,[Y,X]]-\frac{1}{2}[X,[X,Y]]) + O(h^{4}) )},
\end{equation*}
where $[X,Y]=XY-YX$ is the commutator. Therefore, a symmetric splitting (such as DPD-Trotter) guarantees at least second order convergence automatically whereas a nonsymmetric one could only typically gives first order. In what follows we show that the nonsymmetric schemes widely used in DPD simulations provide second order convergence due to cancellations in the error expansions.

In the example of DPD-S1, which can be termed as OBAB and clearly is not symmetric, each interacting pair preserves the invariant distribution $\rho_{\beta}$ (\ref{eqn:Invariant_Distribution_DPD}), i.e.
\begin{equation*}
  \mathcal{L}^{\dag}_{\text{O}_{i,j}} \rho_{\beta} = 0.
\end{equation*}
Thus one can easily verify that the operator $\mathcal{L}^{\dag}_{\text{O}}$ also preserves the invariant distribution
\begin{equation*}
  \mathcal{L}^{\dag}_{\text{O}} \rho_{\beta} = 0,
\end{equation*}
since all the actions of commutators in the BCH expansion of $\mathcal{L}^{\dag}_{\text{O}}$ on the invariant distribution $\rho_{\beta}$ would be zero and therefore all the actions of perturbation operators $\mathcal{L}^{\dag}_{i}$ in Eq. (\ref{eqn:operator_perturbation}) would be zero. Also, it can be easily shown that the Hamiltonian operator
\begin{equation*}
  \mathcal{L}^{\dag}_\text{H} = \mathcal{L}^{\dag}_\text{A} + \mathcal{L}^{\dag}_\text{B}
\end{equation*}
preserves the invariant distribution
\begin{equation*}
  \mathcal{L}^{\dag}_{\text{H}} \rho_{\beta} = 0.
\end{equation*}

Thus, by applying the BCH expansion on the operator of the DPD-S1 scheme (\ref{eqn:DPD_S1_propagator}), we obtain
\begin{equation*}
  \hat{\mathcal{L}}^{\dag}_\text{DPD-S1} = \mathcal{L}^{\dag}_\text{A} + \mathcal{L}^{\dag}_\text{B} + \mathcal{L}^{\dag}_\text{O} + \frac{h}{2} \left( \left[ \mathcal{L}^{\dag}_\text{A}, \mathcal{L}^{\dag}_\text{O} \right] + \left[ \mathcal{L}^{\dag}_\text{B}, \mathcal{L}^{\dag}_\text{O} \right]  \right) + O(h^{2}).
\end{equation*}
Hence,
\begin{equation*}
  \hat{\mathcal{L}}^{\dag}_{1} \rho_{\beta} = \frac{1}{2} \left( \left[ \mathcal{L}^{\dag}_\text{A}, \mathcal{L}^{\dag}_\text{O} \right] + \left[ \mathcal{L}^{\dag}_\text{B}, \mathcal{L}^{\dag}_\text{O} \right]  \right) \rho_{\beta} = \frac{1}{2} \left[ \mathcal{L}^{\dag}_\text{H}, \mathcal{L}^{\dag}_\text{O} \right] \rho_{\beta} = 0,
\end{equation*}
which gives the only solution of the PDE (\ref{eqn:error_analysis_PDE}) to the DPD-S1 scheme
\begin{equation*}
  f_{1} = 0.
\end{equation*}
Given that higher order perturbations in (\ref{eqn:distribution_perturbation}) are not generally equal to zero, we have shown that the nonsymmetric DPD-S1 scheme has second order convergence to its invariant distribution. Similarly, we can also demonstrate that both the Lowe-Andersen and Peters thermostats (in the fashion of BABO) maintain second order convergence to the invariant distribution. Since all the three methods involve a second-order symplectic Verlet method for the deterministic part and other actions only on the Ornstein-Uhlenbeck (OU) process, we may refer to them as generalized Geometric Langevin Algorithms of order two (GLA-2) \cite{Bou-Rabee2010} in the context of stochastic momentum-conserving thermostats. It should be emphasized that the OU process in stochastic momentum-conserving thermostats is pairwise and thus different from the standard setting of Langevin dynamics. Nevertheless, it has been demonstrated that second order is still achieved by the combination of a symplectic method for the deterministic part and an exact solve for the OU process. More discussions regarding the long-time accuracy of the GLA-type methods in Langevin dynamics can be found in \cite{Bou-Rabee2010,Leimkuhler2013c,Abdulle2014}.

\section{Integration Schemes}
\label{Section:Appendix_Schemes}

We list detailed integration steps for each method described in the article here. Verlet neighbor lists \cite{Verlet1967} are used throughout each method as long as possible.

\subsection*{DPD Velocity-Verlet: DPD-VV}

For each particle $i$
\begin{align*}
  \mathbf{p}_{i}^{n+1/2} &= \mathbf{p}_{i}^{n} + \left( h\mathbf{F}^{C}_{i}(\mathbf{q}^{n}) + h\mathbf{F}^{D}_{i}(\mathbf{q}^{n},\mathbf{p}^{n}) + \sqrt{h}\mathbf{F}^{R}_{i}(\mathbf{q}^{n}) \right) /2, \\
  \mathbf{q}_{i}^{n+1} &=  \mathbf{q}_{i}^{n} + hm_{i}^{-1}\mathbf{p}_{i}^{n+1/2}, \\
  \mathbf{p}_{i}^{n+1} &= \mathbf{p}_{i}^{n+1/2} + \left( h\mathbf{F}^{C}_{i}(\mathbf{q}^{n+1}) + h\mathbf{F}^{D}_{i}(\mathbf{q}^{n+1},\mathbf{p}^{n+1/2}) + \sqrt{h}\mathbf{F}^{R}_{i}(\mathbf{q}^{n+1}) \right) /2,
\end{align*}
where $\mathbf{F}^{C}_{i}(\mathbf{q})$, $\mathbf{F}^{D}_{i}(\mathbf{q},\mathbf{p})$ and $\mathbf{F}^{R}_{i}(\mathbf{q})$ are conservative, dissipative and random forces, respectively, in standard DPD system. Note that, at the end of each integration step, the dissipative forces $\mathbf{F}^{D}_{i}(\mathbf{q}^{n+1},\mathbf{p}^{n+1})$ are further updated by using the up-to-date velocities (momenta).

\subsection*{Shardlow's Splitting Method: DPD-S1}

For each interacting pair within cutoff radius ($q_{ij}<r_{c}$)
\begin{align*}
  \mathbf{p}_{i}^{n+1/4} & = \mathbf{p}_{i}^{n} - \gamma\omega^{D}(q_{ij}^{n})(\hat{\mathbf{q}}_{ij}^{n}\cdot \mathbf{v}_{ij}^{n})\hat{\mathbf{q}}_{ij}^{n}h/2 + \sigma\omega^{R}(q_{ij}^{n})\hat{\mathbf{q}}_{ij}^{n} \sqrt{h}\mathrm{R}_{ij}^{n}(t)/2, \\
  \mathbf{p}_{j}^{n+1/4} & = \mathbf{p}_{j}^{n} + \gamma\omega^{D}(q_{ij}^{n})(\hat{\mathbf{q}}_{ij}^{n}\cdot \mathbf{v}_{ij}^{n})\hat{\mathbf{q}}_{ij}^{n}h/2 - \sigma\omega^{R}(q_{ij}^{n})\hat{\mathbf{q}}_{ij}^{n} \sqrt{h}\mathrm{R}_{ij}^{n}(t)/2, \\
  \mathbf{p}_{i}^{n+2/4} & = \mathbf{p}_{i}^{n+1/4} + \sigma\omega^{R}(q_{ij}^{n})\hat{\mathbf{q}}_{ij}^{n} \sqrt{h}\mathrm{R}_{ij}^{n}(t)/2 \\
  & -\frac{\gamma\omega^{D}(q_{ij}^{n}) h}{2(1+\gamma\omega^{D}(q_{ij}^{n}) h)}\left((\hat{\mathbf{q}}_{ij}^{n}\cdot \mathbf{v}_{ij}^{n+1/4})\hat{\mathbf{q}}_{ij}^{n} + \sigma\omega^{R}(q_{ij}^{n})\hat{\mathbf{q}}_{ij}^{n} \sqrt{h}\mathrm{R}_{ij}^{n}(t)\right), \\
  \mathbf{p}_{j}^{n+2/4} & = \mathbf{p}_{j}^{n+1/4} - \sigma\omega^{R}(q_{ij}^{n})\hat{\mathbf{q}}_{ij}^{n} \sqrt{h}\mathrm{R}_{ij}^{n}(t)/2 \\
  & +\frac{\gamma\omega^{D}(q_{ij}^{n}) h}{2(1+\gamma\omega^{D}(q_{ij}^{n}) h)}\left((\hat{\mathbf{q}}_{ij}^{n}\cdot \mathbf{v}_{ij}^{n+1/4})\hat{\mathbf{q}}_{ij}^{n} + \sigma\omega^{R}(q_{ij}^{n})\hat{\mathbf{q}}_{ij}^{n} \sqrt{h}\mathrm{R}_{ij}^{n}(t)\right),
\end{align*}
where $\mathrm{R}_{ij}^{n}(t)$ are normally distributed variables with zero mean and unit variance.

For each particle $i$
\begin{align*}
  \mathbf{p}_{i}^{n+3/4} & = \mathbf{p}_{i}^{n+2/4} + h\mathbf{F}_{i}^{C}(\mathbf{q}^{n})/2, \\
  \mathbf{q}_{i}^{n+1} & = \mathbf{q}_{i}^{n} + hm_{i}^{-1}\mathbf{p}_{i}^{n+3/4}, \\
  \mathbf{p}_{i}^{n+1} & = \mathbf{p}_{i}^{n+3/4} + h\mathbf{F}_{i}^{C}(\mathbf{q}^{n+1})/2.
\end{align*}

\subsection*{DPD-Trotter Scheme: DPD-Trotter}

For each interacting pair within cutoff radius ($q_{ij}<r_{c}$)
\begin{align*}
  \mathbf{p}^{n+1/2}_{i} &= \mathbf{p}^{n}_{i} + h m_{ij}\Delta v_{ij}(\mathbf{q}^{n},\mathbf{p}^{n}) \hat{\mathbf{q}}^{n}_{ij}/2, \\
  \mathbf{p}^{n+1/2}_{j} &= \mathbf{p}^{n}_{j} - h m_{ij}\Delta v_{ij}(\mathbf{q}^{n},\mathbf{p}^{n}) \hat{\mathbf{q}}^{n}_{ij}/2,
\end{align*}
with
\begin{equation*}
    \Delta v_{ij} = \left( \hat{\mathbf{q}}_{ij} \cdot \mathbf{v}_{ij} - \hat{\mathbf{q}}_{ij} \cdot \mathbf{F}_{ij}^{C}(\mathbf{q})/(\tau m_{ij}) \right)( e^{-\tau h} - 1 ) + \sqrt{ k_{B}T(1-e^{-2\tau h})/m_{ij} }\mathrm{R}_{ij}(t),
\end{equation*}
where $\tau=\gamma \omega^{D} / m_{ij}$, $m_{ij}=m_{i}m_{j}/(m_{i}+m_{j})$ and $\mathrm{R}_{ij}(t)$ are normally distributed variables with zero mean and unit variance.

For each particle $i$
\begin{align*}
  \mathbf{q}_{i}^{n+1} & = \mathbf{q}_{i}^{n} + hm_{i}^{-1}\mathbf{p}_{i}^{n+1/2}.
\end{align*}

For each interacting pair within cutoff radius ($q_{ij}<r_{c}$)
\begin{align*}
  \mathbf{p}^{n+1}_{i} &= \mathbf{p}^{n+1/2}_{i} + h m_{ij}\Delta v_{ij}(\mathbf{q}^{n+1},\mathbf{p}^{n+1/2}) \hat{\mathbf{q}}^{n+1}_{ij}/2, \\
  \mathbf{p}^{n+1}_{j} &= \mathbf{p}^{n+1/2}_{j} - h m_{ij}\Delta v_{ij}(\mathbf{q}^{n+1},\mathbf{p}^{n+1/2}) \hat{\mathbf{q}}^{n+1}_{ij}/2.
\end{align*}

\subsection*{Lowe-Andersen Thermostat: LA}

For each particle $i$
\begin{align*}
  \mathbf{p}_{i}^{n+1/3} &= \mathbf{p}_{i}^{n} + h\mathbf{F}_{i}^{C}(\mathbf{q}^{n})/2, \\
  \mathbf{q}_{i}^{n+1} &=  \mathbf{q}_{i}^{n} + hm_{i}^{-1}\mathbf{p}_{i}^{n+1/3}, \\
  \mathbf{p}_{i}^{n+2/3} &= \mathbf{p}_{i}^{n+1/3} + h\mathbf{F}_{i}^{C}(\mathbf{q}^{n+1})/2.
\end{align*}

For each interacting pair within cutoff radius ($q_{ij}<r_{c}$), with probability $P=\Gamma h$
\begin{align*}
  \mathbf{p}^{n+1}_{i} &= \mathbf{p}^{n+2/3}_{i} + \Delta \mathbf{p}_{ij}, \\
  \mathbf{p}^{n+1}_{j} &= \mathbf{p}^{n+2/3}_{j} - \Delta \mathbf{p}_{ij},
\end{align*}
where
\begin{equation*}
  \Delta \mathbf{p}_{ij} = m_{ij} \left( \mathrm{R}_{ij}(t)\sqrt{k_{B}T/m_{ij}} - \hat{\mathbf{q}}^{n+1}_{ij}\cdot \mathbf{v}^{n+2/3}_{ij} \right) \hat{\mathbf{q}}^{n+1}_{ij}.
\end{equation*}

\subsection*{Peters Scheme II: Peters}

For each particle $i$
\begin{align*}
  \mathbf{p}_{i}^{n+1/3} &= \mathbf{p}_{i}^{n} + h\mathbf{F}_{i}^{C}(\mathbf{q}^{n})/2, \\
  \mathbf{q}_{i}^{n+1} &=  \mathbf{q}_{i}^{n} + hm_{i}^{-1}\mathbf{p}_{i}^{n+1/3}, \\
  \mathbf{p}_{i}^{n+2/3} &= \mathbf{p}_{i}^{n+1/3} + h\mathbf{F}_{i}^{C}(\mathbf{q}^{n+1})/2.
\end{align*}

For each interacting pair within cutoff radius ($q_{ij}<r_{c}$)
\begin{align*}
  \mathbf{p}^{n+1}_{i} &= \mathbf{p}^{n+2/3}_{i} + \Delta \mathbf{p}_{ij}, \\
  \mathbf{p}^{n+1}_{j} &= \mathbf{p}^{n+2/3}_{j} - \Delta \mathbf{p}_{ij},
\end{align*}
with
\begin{equation*}
  \Delta \mathbf{p}_{ij} = \left( - \gamma_{ij} (\hat{\mathbf{q}}^{n+1}_{ij}\cdot \mathbf{v}^{n+2/3}_{ij}) h + \sigma_{ij} \sqrt{h}\mathrm{R}_{ij}(t) \right) \hat{\mathbf{q}}^{n+1}_{ij},
\end{equation*}
where
\begin{equation*}
  \gamma_{ij} = \frac{m_{ij}}{h} \left( 1 - \exp\left[ -\frac{\gamma\omega^{D}(q_{ij})h}{m_{ij}} \right] \right), \quad \sigma_{ij} = \frac{k_{B}Tm_{ij}}{h} \left( 1 - \exp\left[ -\frac{2\gamma\omega^{D}(q_{ij})h}{m_{ij}} \right] \right),
\end{equation*}
where $\gamma$ and $\sigma$ are the same dissipation and random coefficients respectively as in standard DPD system.

\subsection*{Nos\'{e}-Hoover-Lowe-Andersen Thermostat: NHLA}

For each particle $i$
\begin{align*}
  \mathbf{q}_{i}^{n+1} &= \mathbf{q}_{i}^{n} + h\mathbf{v}_{i}^{n} + h^{2}\mathbf{F}_{i}^{C}(\mathbf{q}^{n})/2, \\
  \mathbf{v}_{i}^{n+1/4} &=  \mathbf{v}_{i}^{n} + h\mathbf{F}_{i}^{C}(\mathbf{q}^{n})/2.
\end{align*}

For ($1-P$) fraction interacting pairs within cutoff radius ($q_{ij}<r_{c}$)
\begin{align*}
  \mathbf{F}_{i}^{D}(\mathbf{q}^{n+1},\mathbf{p}^{n+1/4}) &= \mathbf{F}_{i}^{D}(\mathbf{q}^{n},\mathbf{p}^{n}) + \mathbf{F}_{ij}^{D}(\mathbf{q}^{n+1},\mathbf{p}^{n+1/4}), \\
  \mathbf{F}_{j}^{D}(\mathbf{q}^{n+1},\mathbf{p}^{n+1/4}) &= \mathbf{F}_{j}^{D}(\mathbf{q}^{n},\mathbf{p}^{n}) - \mathbf{F}_{ij}^{D}(\mathbf{q}^{n+1},\mathbf{p}^{n+1/4}),
\end{align*}
where
\begin{equation*}
   \mathbf{F}_{ij}^{D}(\mathbf{q}^{n+1},\mathbf{p}^{n+1/4}) = \alpha \omega^{R}(q_{ij}) (\hat{\mathbf{q}}^{n+1}_{ij}\cdot \mathbf{v}^{n+1/4}_{ij})\hat{\mathbf{q}}^{n+1}_{ij},
\end{equation*}
where $\alpha$ is a coupling parameter chosen as $0.9/(\rho h)$, and $\rho$ is the particle density.

For each particle $i$
\begin{align*}
   \mathbf{v}_{i}^{n+2/4} &=  \mathbf{v}_{i}^{n+1/4} + h\mathbf{F}_{i}^{C}(\mathbf{q}^{n+1})/2, \\
   \mathbf{p}^{n+3/4}_{i} &= \mathbf{p}^{n+2/4}_{i} + h(1-\tilde{T}_{k}/T_{0})\mathbf{F}_{i}^{D}(\mathbf{q}^{n+1},\mathbf{p}^{n+1/4}),
\end{align*}
where $\tilde{T}_{k}$ is the momentary kinetic temperature and $T_{0}$ is the desired temperature.

For the remaining ($P$) fraction interacting pairs within cutoff radius ($q_{ij}<r_{c}$)
\begin{align*}
  \mathbf{p}^{n+1}_{i} &= \mathbf{p}^{n+3/4}_{i} + \Delta \mathbf{p}_{ij}, \\
  \mathbf{p}^{n+1}_{j} &= \mathbf{p}^{n+3/4}_{j} - \Delta \mathbf{p}_{ij},
\end{align*}
where
\begin{equation*}
  \Delta \mathbf{p}_{ij} = m_{ij} \left( \mathrm{R}_{ij}(t)\sqrt{k_{B}T/m_{ij}} - \hat{\mathbf{q}}^{n+1}_{ij}\cdot \mathbf{v}^{n+3/4}_{ij} \right) \hat{\mathbf{q}}^{n+1}_{ij}.
\end{equation*}

\subsection*{Pairwise Nos\'{e}-Hoover Thermostat: PNH}

For each particle $i$
\begin{align*}
  \mathbf{p}_{i}^{n+1/2} &= \mathbf{p}_{i}^{n} + h \left( \mathbf{F}^{C}_{i}(\mathbf{q}^{n}) - \xi^{n} \mathbf{V}_{i}(\mathbf{q}^{n},\mathbf{p}^{n-1/2}) \right)/2, \\
  \xi^{n+1/2} &=  \xi^{n} + hG(\mathbf{q}^{n},\mathbf{p}^{n-1/2})/2, \\
  \mathbf{q}_{i}^{n+1} &=  \mathbf{q}_{i}^{n} + hm_{i}^{-1}\mathbf{p}_{i}^{n+1/2} , \\
  \mathbf{p}_{i}^{n+1} &= \mathbf{p}_{i}^{n+1/2} + h \left( \mathbf{F}^{C}_{i}(\mathbf{q}^{n+1}) - \xi^{n+1/2} \mathbf{V}_{i}(\mathbf{q}^{n+1},\mathbf{p}^{n+1/2}) \right)/2, \\
  \xi^{n+1} &=  \xi^{n+1/2} + hG(\mathbf{q}^{n+1},\mathbf{p}^{n+1/2})/2,
\end{align*}
where
\begin{align*}
  \mathbf{V}_{i}(\mathbf{q},\mathbf{p})   &= \sum_{j\neq i}\omega^{D}(q_{ij})(\hat{\mathbf{q}}_{ij}\cdot \mathbf{v}_{ij})\hat{\mathbf{q}}_{ij}, \\
  G(\mathbf{q},\mathbf{p}) &= {\mu}^{-1}\sum_{i}\sum_{j>i}\omega^{D}(q_{ij}) \left[ \left( \mathbf{v}_{ij}\cdot\hat{\mathbf{q}}_{ij} \right)^{2} - k_{B}T/m_{ij}  \right].
\end{align*}

\subsection*{Symmetric Pairwise Nos\'{e}-Hoover-Langevin Thermostat: PNHL-S}

For each particle $i$
\begin{align*}
  \mathbf{q}_{i}^{n+1/2} &=  \mathbf{q}_{i}^{n} + hm_{i}^{-1}\mathbf{p}_{i}^{n}/2, \\
  \mathbf{p}_{i}^{n+1/4} &= \mathbf{p}_{i}^{n} + h\mathbf{F}_{i}^{C}(\mathbf{q}^{n+1/2})/2.
\end{align*}

For each interacting pair within cutoff radius ($q_{ij}<r_{c}$)
\begin{align*}
  \mathbf{p}^{n+2/4}_{i} &= \mathbf{p}^{n+1/4}_{i} + h m_{ij}\Delta v_{ij}(\mathbf{q}^{n+1/2},\mathbf{p}^{n+1/4},\xi^{n}) \hat{\mathbf{q}}^{n+1/2}_{ij}/2, \\
  \mathbf{p}^{n+2/4}_{j} &= \mathbf{p}^{n+1/4}_{j} - h m_{ij}\Delta v_{ij}(\mathbf{q}^{n+1/2},\mathbf{p}^{n+1/4},\xi^{n}) \hat{\mathbf{q}}^{n+1/2}_{ij}/2,
\end{align*}
where
\begin{equation*}
    \Delta v_{ij} = \left( \hat{\mathbf{q}}_{ij} \cdot \mathbf{v}_{ij} \right) \left( \exp(-\xi \omega^{D} h / m_{ij} ) - 1 \right).
\end{equation*}

For additional variable $\xi$
\begin{align*}
  \xi^{n+1/3} &=  \xi^{n} + hG(\mathbf{q}^{n+1/2},\mathbf{p}^{n+2/4})/2, \\
  \xi^{n+2/3} &= e^{-\gamma^{\ast} h}\xi^{n+1/3} + \sqrt{ k_{B}T(1-e^{-2\gamma^{\ast} h})/\mu }\mathrm{R}(t), \\
  \xi^{n+1} &=  \xi^{n+2/3} + hG(\mathbf{q}^{n+1/2},\mathbf{p}^{n+2/4})/2,
\end{align*}
where $\mathrm{R}(t)$ are normally distributed variables with zero mean and unit variance.

For each interacting pair within cutoff radius ($q_{ij}<r_{c}$)
\begin{align*}
  \mathbf{p}^{n+3/4}_{i} &= \mathbf{p}^{n+2/4}_{i} + h m_{ij}\Delta v_{ij}(\mathbf{q}^{n+1/2},\mathbf{p}^{n+2/4},\xi^{n+1}) \hat{\mathbf{q}}^{n+1/2}_{ij}/2, \\
  \mathbf{p}^{n+3/4}_{j} &= \mathbf{p}^{n+2/4}_{j} - h m_{ij}\Delta v_{ij}(\mathbf{q}^{n+1/2},\mathbf{p}^{n+2/4},\xi^{n+1}) \hat{\mathbf{q}}^{n+1/2}_{ij}/2.
\end{align*}

For each particle $i$
\begin{align*}
  \mathbf{p}_{i}^{n+1} &= \mathbf{p}_{i}^{n+3/4} + h\mathbf{F}_{i}^{C}(\mathbf{q}^{n+1/2})/2, \\
  \mathbf{q}_{i}^{n+1} &=  \mathbf{q}_{i}^{n+1/2} + hm_{i}^{-1}\mathbf{p}_{i}^{n+1}/2.
\end{align*}

\subsection*{Nonsymmetric Pairwise Nos\'{e}-Hoover-Langevin Thermostat: PNHL-N}

For each particle $i$
\begin{align*}
  \mathbf{q}_{i}^{n+1/2} &=  \mathbf{q}_{i}^{n} + hm_{i}^{-1}\mathbf{p}_{i}^{n}/2, \\
  \mathbf{p}_{i}^{n+1/4} &= \mathbf{p}_{i}^{n} + h\mathbf{F}_{i}^{C}(\mathbf{q}^{n+1/2})/2.
\end{align*}

For each interacting pair within cutoff radius ($q_{ij}<r_{c}$)
\begin{align*}
  \mathbf{p}^{n+2/4}_{i} &= \mathbf{p}^{n+1/4}_{i} + h m_{ij}\Delta v_{ij}(\mathbf{q}^{n+1/2},\mathbf{p}^{n+1/4},\xi^{n}) \hat{\mathbf{q}}^{n+1/2}_{ij}/2, \\
  \mathbf{p}^{n+2/4}_{j} &= \mathbf{p}^{n+1/4}_{j} - h m_{ij}\Delta v_{ij}(\mathbf{q}^{n+1/2},\mathbf{p}^{n+1/4},\xi^{n}) \hat{\mathbf{q}}^{n+1/2}_{ij}/2,
\end{align*}
where
\begin{equation*}
    \Delta v_{ij} = \left( \hat{\mathbf{q}}_{ij} \cdot \mathbf{v}_{ij} \right) \left( \exp(-\xi \omega^{D} h / m_{ij} ) - 1 \right).
\end{equation*}

For additional variable $\xi$
\begin{align*}
  \xi^{n+1/3} &=  \xi^{n} + hG(\mathbf{q}^{n+1/2},\mathbf{p}^{n+2/4})/2, \\
  \xi^{n+2/3} &= e^{-\gamma^{\ast} h}\xi^{n+1/3} + \sqrt{ k_{B}T(1-e^{-2\gamma^{\ast} h})/\mu }\mathrm{R}(t), \\
  \xi^{n+1} &=  \xi^{n+2/3} + hG(\mathbf{q}^{n+1/2},\mathbf{p}^{n+2/4})/2.
\end{align*}

For each interacting pair within cutoff radius ($q_{ij}<r_{c}$)
\begin{align*}
  \mathbf{p}^{n+3/4}_{i} &= \mathbf{p}^{n+2/4}_{i} + h m_{ij}\Delta v_{ij}(\mathbf{q}^{n+1/2},\mathbf{p}^{n+2/4},\xi^{n+1}) \hat{\mathbf{q}}^{n+1/2}_{ij}/2, \\
  \mathbf{p}^{n+3/4}_{j} &= \mathbf{p}^{n+2/4}_{j} - h m_{ij}\Delta v_{ij}(\mathbf{q}^{n+1/2},\mathbf{p}^{n+2/4},\xi^{n+1}) \hat{\mathbf{q}}^{n+1/2}_{ij}/2.
\end{align*}

For each particle $i$
\begin{align*}
  \mathbf{q}_{i}^{n+1} &=  \mathbf{q}_{i}^{n+1/2} + hm_{i}^{-1}\mathbf{p}_{i}^{n+3/4}/2, \\
  \mathbf{p}_{i}^{n+1} &= \mathbf{p}_{i}^{n+3/4} + h\mathbf{F}_{i}^{C}(\mathbf{q}^{n+1})/2.
\end{align*}

\end{appendices}




\bibliographystyle{is-abbrv}

\bibliography{refs}

\end{document}